\documentclass[onecolumn]{aa}  
\usepackage{graphicx}
\usepackage{txfonts}
\begin{document}
   \title{Precise Modeling of the Exoplanet Host Star and CoRoT Main Target HD 52265}

%   \subtitle{I. Overviewing the $\kappa$-mechanism}

   \author{M. E. Escobar
          \inst{1}, S. Th\'eado\inst{1}, S. Vauclair\inst{1,2}, J. Ballot\inst{1}, S. Charpinet\inst{1}, N. Dolez\inst{1}, 
          A. Hui-Bon-Hoa\inst{1}, G. Vauclair\inst{1}, L. Gizon\inst{3,4}, S. Mathur\inst{5},  P.O. Quirion\inst{6}, T. Stahn\inst{4}
          }

\institute{Institut de Recherche en Astrophysique et Plan\'etologie, Observatoire Midi-Pyr\'en\'ees, CNRS,
              Universit\'e Paul Sabatier, 14 Avenue Edouard Belin, 31400 Toulouse, France.
             %\email{mescobar@ast.obs-mip.fr}
           \and
           Institut universitaire de France, 103 boulevard Saint Michel, 75005 Paris, France
           \and
           Max-Planck-Institut f\"ur Sonnensystemforschung, Max-Planck-Str. 2, 37191 Katlenburg-Lindau, Germany
           \and
           Institut f\"ur Astrophysik, Georg-August-Universit\"at G\"ottingen, Friedrich-Hundt-Platz 1, 37077 G\"ottingen Germany
            \and
            High Altitude Observatory, NCAR, P.O. Box 3000, Boulder, CO 80307, USA 
            \and
	    Canadian Space Agency, 6767 Boulevard de l'Aéroport, Saint-Hubert, QC, J3Y 8Y9, Canada\\
            \email{mescobar@irap.omp.eu}
             }

   \date{Received XXXX, XXXX; accepted XXXX, XXXX}

% \abstract{}{}{}{}{} 
% 5 {} token are mandatory
 
  \abstract
   {}
   {This paper presents a detailed and precise study of the characteristics of the Exoplanet Host Star and CoRoT main target HD 52265, as derived from asteroseismic studies. The results are compared with previous estimates, with a comprehensive summary and discussion.}
   {The basic method is similar to that previously used by the Toulouse group for solar-type stars. Models are computed with various initial chemical compositions and the computed p-mode frequencies are compared with the observed ones. All models include atomic diffusion and the importance of radiative accelerations is discussed. Several tests are used, including the usual frequency combinations and the fits of the \'echelle diagrams. The possible surface effects are introduced and discussed. Automatic codes are also used to find the best model for this star (SEEK, AMP) and their results are compared with that obtained with the detailed method.}
   {We find precise results for the mass, radius and age of this star, as well as its effective temperature and luminosity. We also give an estimate of the initial helium abundance. These results are important for the characterization of the star-planet system.}
   {}


\authorrunning{M. E. Escobar et al.}
\titlerunning{HD 52265}

   \maketitle

%________________________________________________________________

\section{Introduction}

In the last few years the number of observed exoplanets have increased dramatically, owing to 
missions like CoRoT (Baglin et al. \cite{baglin06}) and \textit{Kepler} (Koch et al. \cite{koch10}), as well as many other ground based and space missions 
(see the exoplanet encyclopedia for a complete summary \footnote[1]{http://exoplanet.eu/index.php}). 
Precise characterization of exoplanet host stars become more and more important in the framework of the detailed studies of the observed planetary systems. Constraints on the parameters and internal structure of the star can be obtained by comparing models with photometric and 
spectroscopic observations (Southworth \cite{southworth11}; Basu et al. \cite{basu12}), but the best precision is obtained from asteroseismology, when the stellar 
oscillations may be observed and analysed. This was the case, for example, 
for the exoplanet host stars (hereafter EHS) $\iota$ Hor (Vauclair et al. \cite{vauclair08}) and $\mu$ Arae
(Soriano \& Vauclair \cite{soriano10}), both observed with HARPS, 
as well as the EHS HAT-P-7, HAT-P-11 and TrES-2 (Christensen-Dalsgaard et al. \cite{jcd10}), 
observed with \textit{Kepler}. 

Among the EHS, the star HD 52265 is one of the most precisely observed for asteroseismology, as it was the only EHS observed as a main target by CoRoT.
This G0V metal-rich main sequence star has an orbiting jupiter-mass planet at 0.5 AU with a period of
119 days (Naef et al. \cite{naef01}; Butler et al. \cite{butler00}). 
It was continuously observed between December 13, 2008 and March 3, 2009, that is 117 consecutive days. As a result, 31 p-mode frequencies were reported, 
between 1500--2550 $\mu$Hz, corresponding to $\ell$ = 0, 1 and 2 (Ballot et al. \cite{ballot11}). From this analysis, a large separation of 
$<\Delta \nu>$ = 98.4 $\pm$ 0.1 $\mu$Hz and a small separation of $<\delta \nu_{02}>$=8.1 $\pm$ 0.2 $\mu$Hz were found, 
and a complete asteroseismic analysis including mode lifetimes was presented. An extensive study of the seismic rotation of HD 52265 was performed by Stahn \cite{stahn11} and Gizon et al. \cite{gizon12}.

Spectroscopic observations of this star were done by several groups, who gave different values of the observed triplet ([Fe/H], log g, Teff). 
Their results are given in Table 1. Some of these groups also observed lines of other elements, and gave detailed relative abundances. The results show an overall surmetallicity in this star, similar for most heavy elements, with a small dispersion.
The Hipparcos parallax is 34.54 $\pm$ 0.40 mas (van Leeuwen \cite{leeuwen07}), 
which leads to a luminosity value log L/L$_{\sun}=0.29 \pm 0.05$.
A spectroscopic follow up was also done with the Narval spectropolarimeter installed on the Bernard Lyot telescope
at Pic du Midi Observatory (France) during December 2008 and January 2009, i.e., during CoRoT observations.
No magnetic signature was observed.

Preliminary modeling of this star, using spectroscopic constraints, was done by Soriano et al. (\cite{soriano07}), as a preparation to CoRoT observations. 
Evolutionary tracks were computed using the Toulouse-Geneva evolution code (TGEC). 
According to the spectroscopic constraints, eight models with masses between 1.18 and 1.30 M$_{\sun}$ and metallicities ranging 
from 0.19 to 0.27 were chosen for further analysis, and adiabatic p-modes frequencies were computed. Echelle diagrams for each selected model
were presented as well as their corresponding large and small separations. A large separation around  $\sim$ 100 $\mu$Hz 
was predicted from these models except for the Takeda et al. (\cite{takeda05}) values, which corresponded to a smaller large 
separation (around $\sim$ 75 $\mu$Hz).

The detailed CoRoT observations allow going further in this analysis, using the precise seismic results. First of all, the 
Takeda et al. (\cite{takeda05}) values,which correspond to a more evolved star, are excluded. 
Now we present a complete asteroseismic analysis for HD 52265, and give precise results on the stellar parameters.

The method and models used for the asteroseismic comparisons with observations are described in section 2. All the models, 
as described in section 2.1, include element gravitational settling. The seismic tests are discussed in section 2.2 and we 
discuss the influence of radiative accelerations on heavy elements in section 2.3. The results are given in section 3. 
Section 3.1 is devoted to the results obtained without taking surface effects into account. We test the influence of 
varying the initial chemical composition and the introduction or not of radiative levitation on heavy 
elements. An analysis of surface effects and the consequences on the results is given in section 3.2. 
A first discussion of the results is given in section 3.3. Finally, in section 4 the results obtained using automatic 
codes to find the best models for this star from seismology (SEEK and AMP) are presented and compared with the previously 
obtained solutions. Summary and discussion are given in section 5.

%%%%%%%%%%%%%%%%%%%%%%%%%%%%%%%%%% TABLA 1 PREVIOUS SPECTROSCOPIC STUDIES %%%%%%%%%%%%%%%%%%%%%%%%%%%%%%%

\begin{table}[h]
\caption{Summary of previous spectroscopic studies of HD 52265}
\label{prevspec}
\centering
\begin{tabular}{c c c l}     % 4 columns
\hline\hline

[Fe/H] & T$_\mathrm{eff}$ &  log $g$ &  Reference \\
\hline 
0.27 $\pm$ 0.02 & 6162 $\pm$ 22 & 4.29 $\pm$ 0.04 & Gonzalez et al. 2001  \\
0.23 $\pm$ 0.05 & 6103 $\pm$ 52 & 4.28 $\pm$ 0.12 & Santos et al. 2004    \\
0.19 $\pm$ 0.03 & 6069 $\pm$ 15 & 4.12 $\pm$ 0.09 & Takeda et al. 2005  \\
0.19 $\pm$ 0.03 & 6076 $\pm$ 44 & 4.26 $\pm$ 0.06 & Fischer \& Valenti 2005 \\
0.24 $\pm$ 0.02 & 6179 $\pm$ 18 & 4.36 $\pm$ 0.03 & Gillon \& Magain 2006  \\
0.19 $\pm$ 0.05 & 6100 $\pm$ 60 & 4.35 $\pm$ 0.09 & Ballot et al. 2011     \\

\hline
\end{tabular}
\end{table}

%%%%%%%%%%%%%%%%%%%%%%%%%%%%%%%%%%%%%%%%%%%%%%%%%%%%%%%%%%%%%%%%%%%%%%%%%%%%%%%%%%%%%%%%%%%%%%%%%%%%%%%%%%%%%%%%%

%________________________________________________________________________________________________________

\section{Computations with TGEC}

\subsection{Stellar models}

Stellar models were computed using the TGEC code, (Hui-Bon-Hoa \cite{hui08}; Th\'eado et al. \cite{theado12}), with the OPAL
equation of state and opacities (Rogers \& Nayfonov \cite{rogers02}; Iglesias \& Rogers \cite{iglesias96}), and the NACRE nuclear reaction 
rates (Angulo et al. \cite{angulo99}). Convection was treated
using the mixing length theory. For all models, the mixing length parameter was adjusted to that of the solar case, i.e. $\alpha = 1.8$ 
without overshooting and without extra-mixing. Gravitational settling of helium and metals was included using the Paquette prescription 
(Paquette et al. \cite{paquette86}; Michaud et al. \cite{michaud04}). Radiative accelerations of metals were also introduced using the 
SVP method (Single Valued Parameters approximation, see Alecian \& LeBlanc \cite{alecian02}, LeBlanc \& Alecian \cite{leblanc04} and 
Th\'eado et al. \cite{theado09}). As most stellar evolution codes neglect these radiative accelerations, we analysed the effects on 
the seismic results of introducing them or not.
 
Evolutionary tracks were computed for two metallicity values and two different initial helium abundances. The metallicity values 
were chosen as [Fe/H] = 0.23 and 0.30, so that after diffusion the final model value lies inside the observed range. Here [Fe/H] 
represents the global overmetallicity with respect to the Sun, defined as $[log(Z/X)_*-log(Z/X)_{\sun}]$, where $Z$ and $X$ are 
computed at the stellar surface. Considering the small dispersion of the detailed abundances, this value may be compared with the 
observed [Fe/H]. A discussion of the computed detailed abundance variations is given in section 2.3. The initial helium values are 
labeled as Y$_{\sun}$ and Y$_{G}$, where  Y$_{\sun}$ is the solar helium value taken from Grevesse \& Noels \cite{grevesse93} and 
Y$_{G}$ is a helium abundance which increases with Z as expected if the local medium follows the general trend observed for the
chemical evolution of galaxies (C.f. Izotov \& Thuan \cite{izotov04} and \cite{izotov10}).

Adiabatic oscillation frequencies were computed for many models along the evolutionary tracks 
using the PULSE code (Brassard \& Charpinet \cite{brassard08}). We computed these frequencies for degrees
$\ell$ = 0 to $\ell$ = 3. For comparisons with the observations (seismic tests) we used the same frequency interval for the computed 
as for the observed frequencies for consistency, i.e. between 1500 and 2550 $\mu$Hz, as discussed below.

\subsection{Seismic Tests}
A well known characteristic of p modes in spherical stars is that modes of the same degree with successive radial order $n$
are nearly equally spaced in frequency (e.g. Tassoul \cite{tassoul80}). The large separation is defined as:

\begin{equation}
      \Delta \nu _{n,\ell} = \nu_{\mathrm{n+1,\ell}} - \nu_{\mathrm{n,\ell}}
\end{equation}

In real stars, this large separation slightly varies with frequency, so that an average value has to be used for comparisons between models and observations. One has to be careful to use the same frequency range in both cases to do the comparisons. Taking this into account, the fit between the computed and measured large separations is the first step in the process of comparisons. The large separation gives access to the stellar average density (Ulrich \cite{ulrich86}; White et al. \cite{white11}).

A second characteristic of p modes is that the difference between $(n,\ell)$ modes and $(n-1, \ell+2)$ ones varies very slowly with frequency. The small separations are defined as:

\begin{equation}
      \delta \nu _{n,\ell} = \nu_{\mathrm{n,\ell}} - \nu_{\mathrm{n-1,\ell+2}}
\end{equation}

These small separations are most sensitive to the stellar core and may give information on the extension of the convective core in some stars 
(Tassoul \cite{tassoul80}; Roxburg \& Vorontsov \cite{roxburgh94}; Gough \cite{gough86}; Soriano \& Vauclair \cite{soriano08}).

Provided that the stellar chemical composition is precisely known, the knowledge of both the large and the small separations, which may be plotted in the so-called C-D diagrams, gives good constraints on the stellar parameters (Christensen-Dalsgaard \cite{jcd84}; White et al. \cite{white11}). However, whereas the stellar metallicity can be precisely derived from spectroscopy, the helium content of solar-type stars is not directly known from observations. This ignorance leads to important uncertainties on the evolutionary tracks, and thus on the derived stellar parameters. 

We analysed these uncertainties in detail by computing models with various chemical compositions. For each stellar evolutionary track that we computed, we first searched for the model which had an average large separation of $<\Delta \nu>$ = 98.4 $\pm$ 0.1 $\mu$Hz. As the large separation continously decreases when the star evolves along the main sequence, one model only is found with the observed value (within the uncertainties). Then, for each set of computations done with a given initial chemical composition ([Fe/H] and Y), we looked for the model which best fitted the small separations observed between modes of $\ell$ =2 and 0, $<\delta \nu_{02}>$=8.1 $\pm$ 0.2 $\mu$Hz. We further proceed with detailed comparisons of observed and computed \'echelle diagrams.

The final comparison between the models and the seismic observations needs taking into account the surface effects induced by the waves behavior in the outer stellar layers.  We computed the frequency shift induced by such effects, using the recipe proposed by Kjeldsen et al. (\cite{kjeldsen08}). In this case, the large separations are modified, as discussed in section 3.2, which leads to corrections in the results.

%%%%%%%%%%%%%%%%%%%%%%%%%%%%%%%  FIGURE ?? - EVOLUTIONARY TRACKS ALPHA 1.8 %%%%%%%%%%%%%%%%%%%%%%%%%%%%%%%%%%%%%%%%%%%%%%%%%%%%%%%%%

   \begin{figure*}
   \centering
   \includegraphics[width=0.48\textwidth]{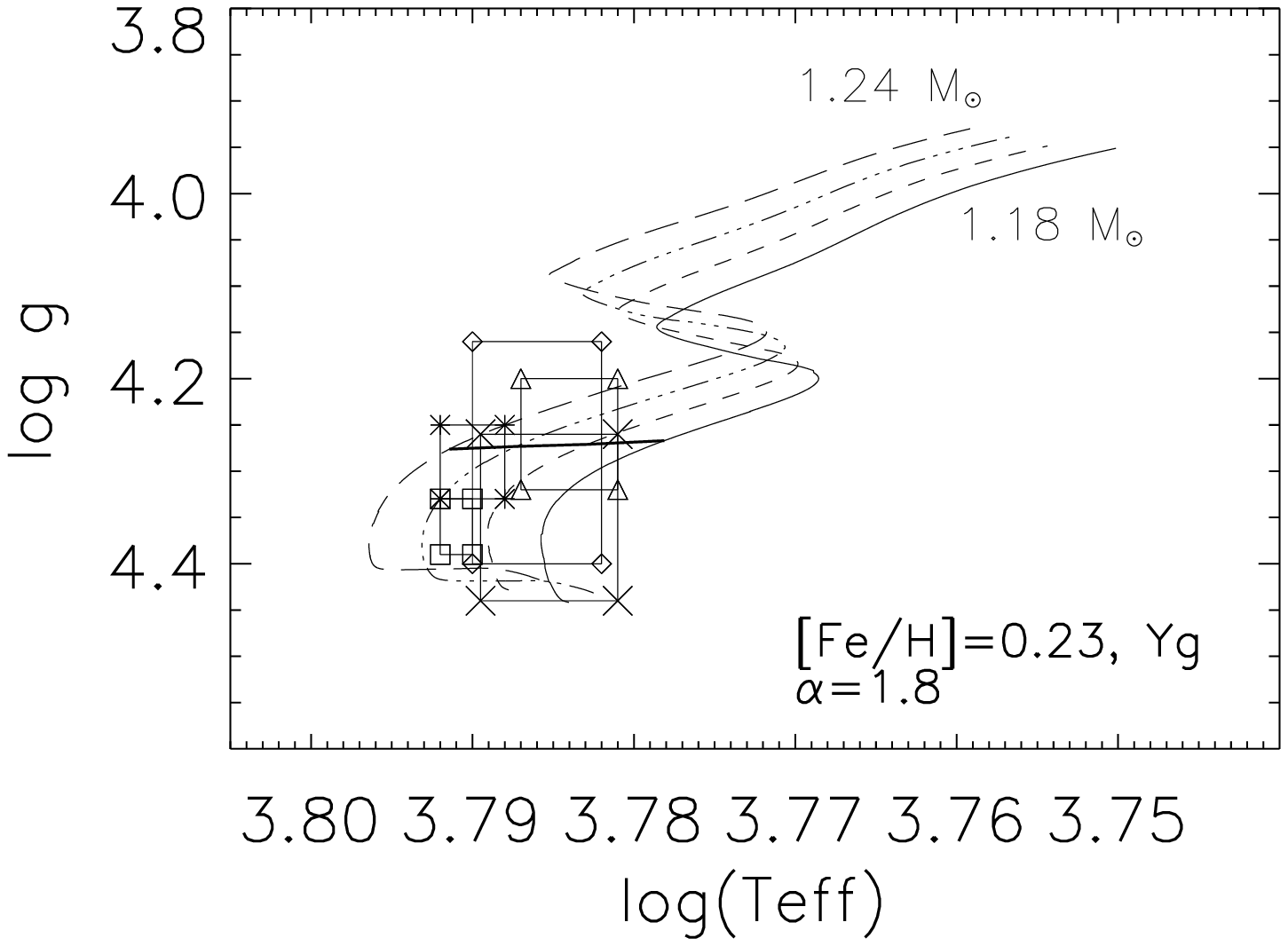}
   \includegraphics[width=0.48\textwidth]{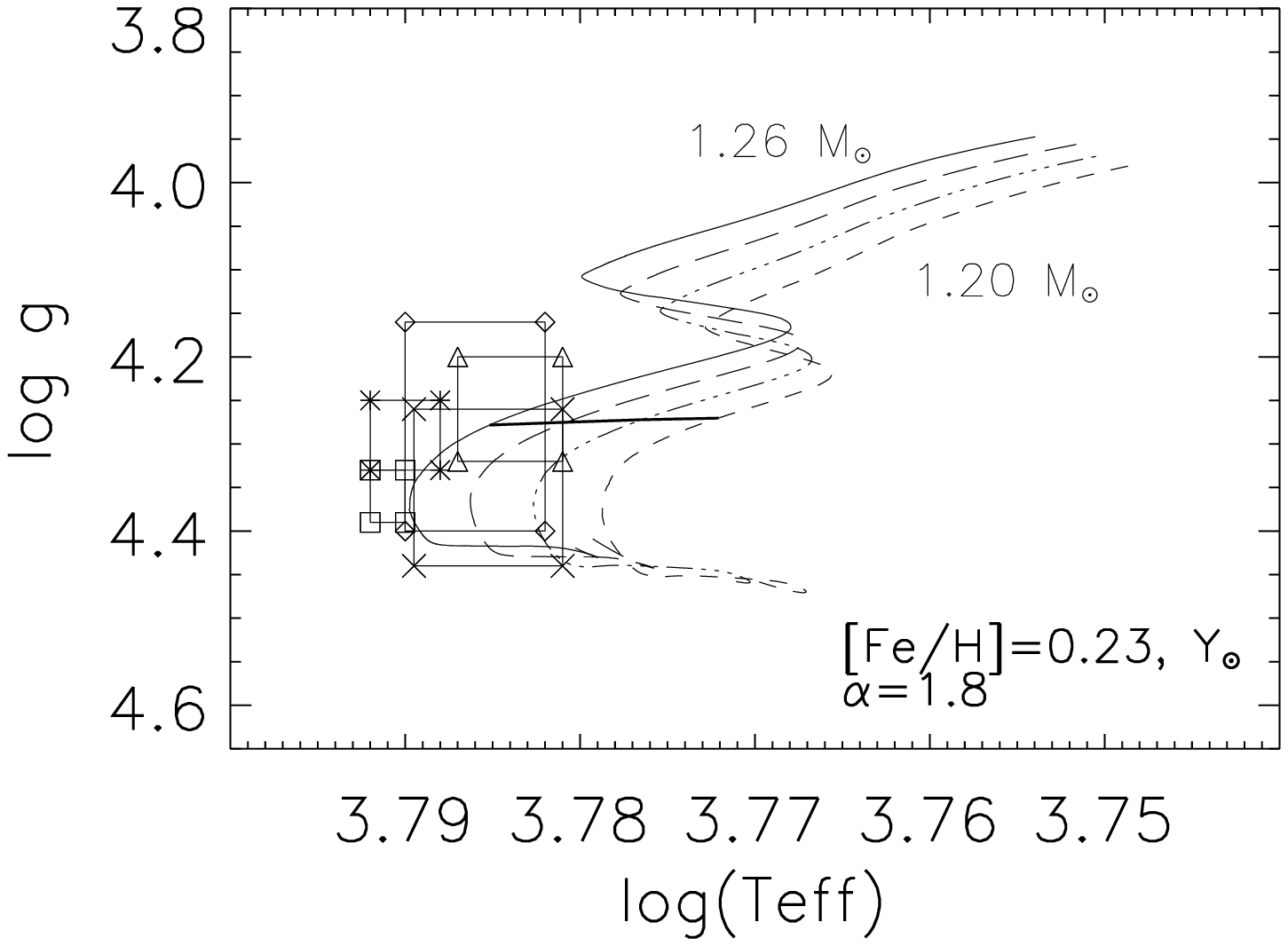}
   \includegraphics[width=0.48\textwidth]{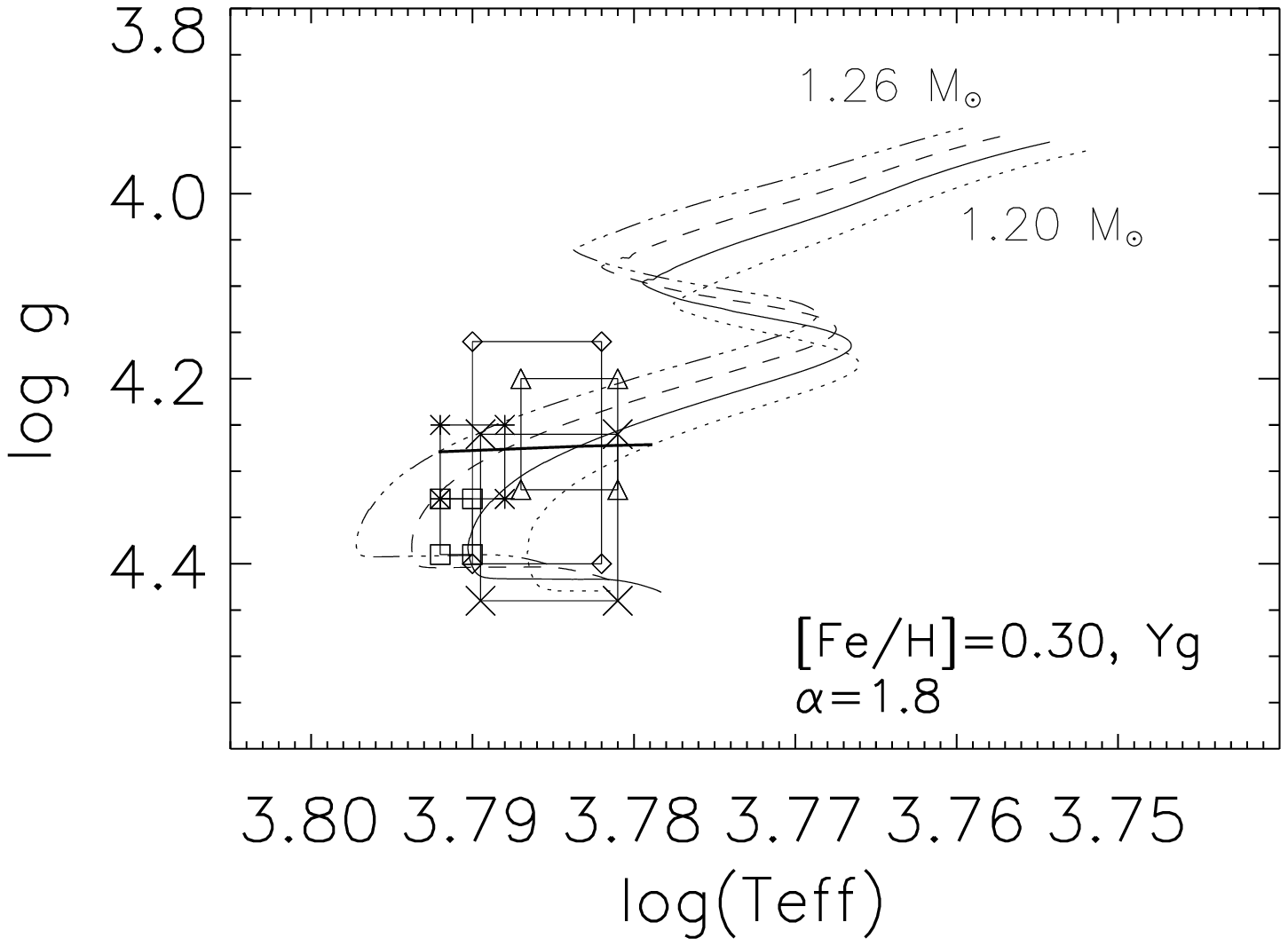}
   \includegraphics[width=0.48\textwidth]{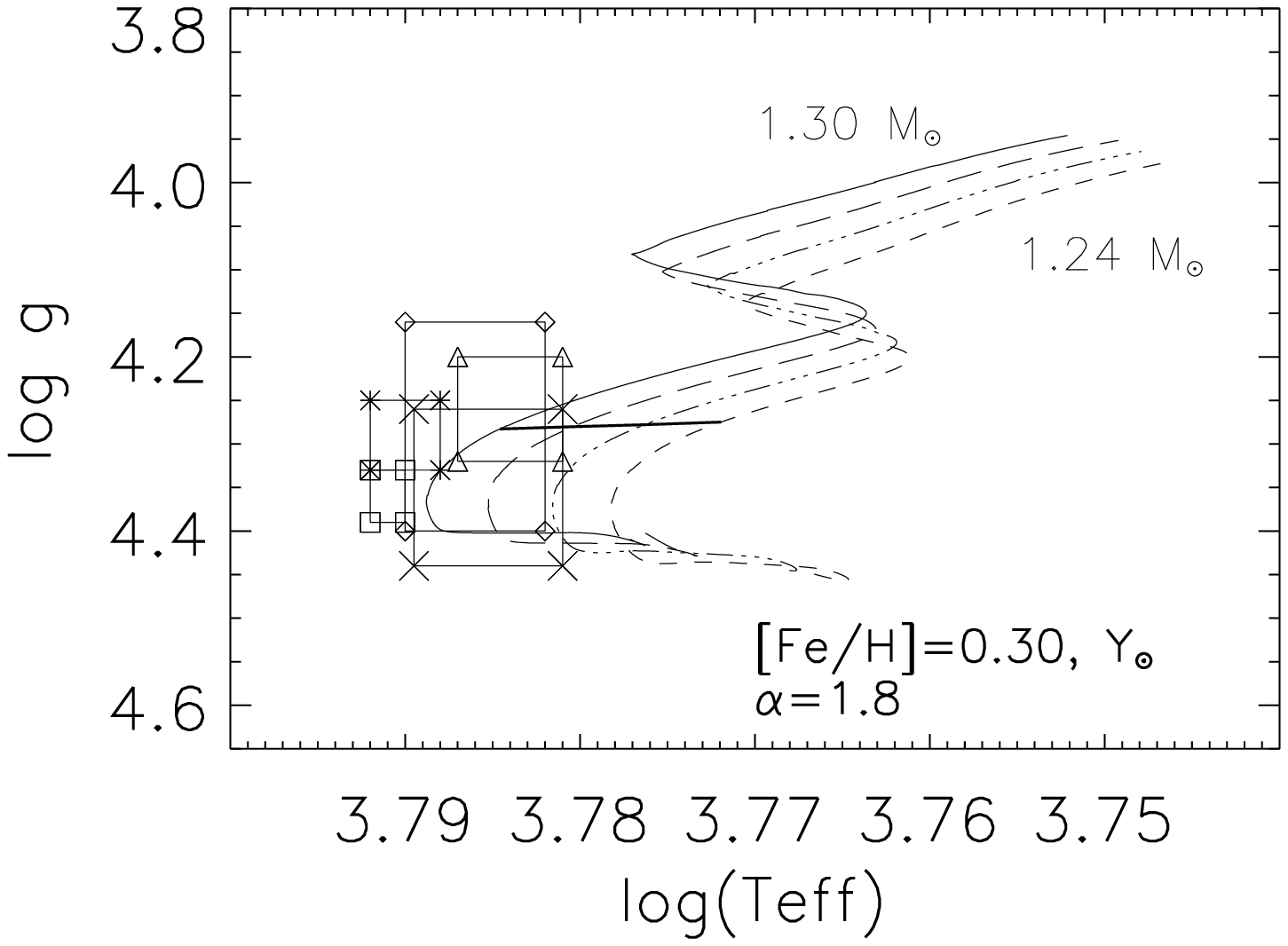}
     \caption{Evolutionary tracks in the log $g$ vs log $T_\mathrm{eff}$ plane for the various sets of metallicity and helium abundance, for $\alpha=1.8$ (see text for details).
     The symbols indicate the error boxes of Gonzalez et al. (\cite{gonzalez01}) (\textit{asterisks}), Santos et al. (\cite{santos04}) (\textit{diamonds}),
     Gillon \& Magain (\cite{gillon06}) (\textit{squares}), Fisher \& Valenti (\cite{fisher05}) (\textit{triangles}), and Ballot et al. (\cite{ballot11}) (\textit{crosses}). 
     The straight thick line represents the iso-$<\Delta \nu>$ line, with $<\Delta \nu>$=98.4 $\mu$Hz.}
              \label{logtracks1}
    \end{figure*}

%%%%%%%%%%%%%%%%%%%%%%%%%%%%%%%%%%%%%%%%%%%%%%%%%%%%%%%%%%%%%%%%%%%%%%%%%%%%%%%%%%%%%%%%%%%%%%%%%%%%%%%%%%%%%%%%%%%%%%%%%%%%%%

%%%%%%%%%%%%%%%%%%%%%%%%%%%%%%% FIGURE EVOLUTIONARY TRACKS ALPHA 1.8 %%%%%%%%%%%%%%%%%%%%%%%%%%%%%%%%%%%%%%%%%%%%%%%%%%%%%%%%%

   \begin{figure*}
   \centering
   \includegraphics[width=0.48\textwidth]{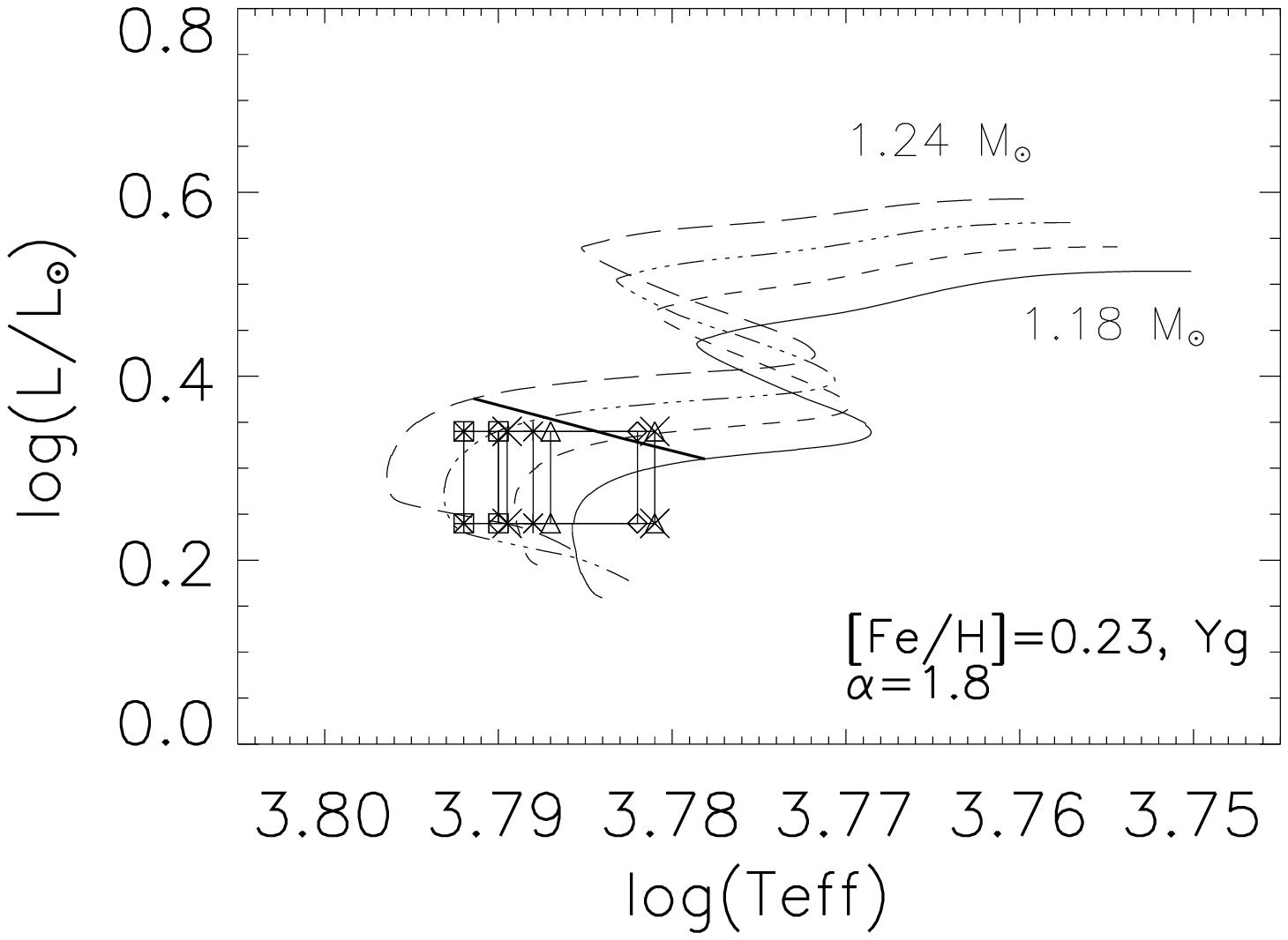}
   \includegraphics[width=0.48\textwidth]{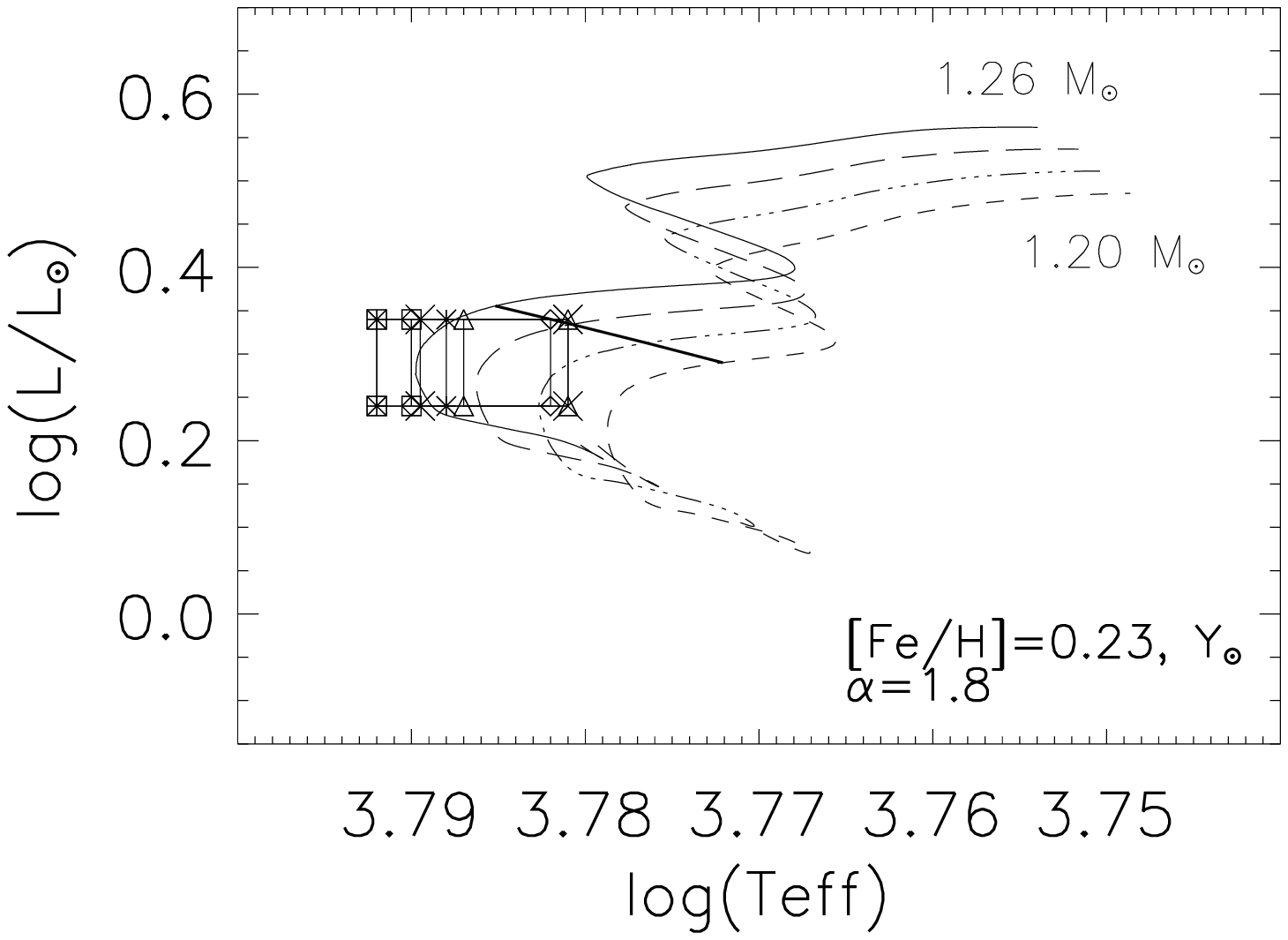}
   \includegraphics[width=0.48\textwidth]{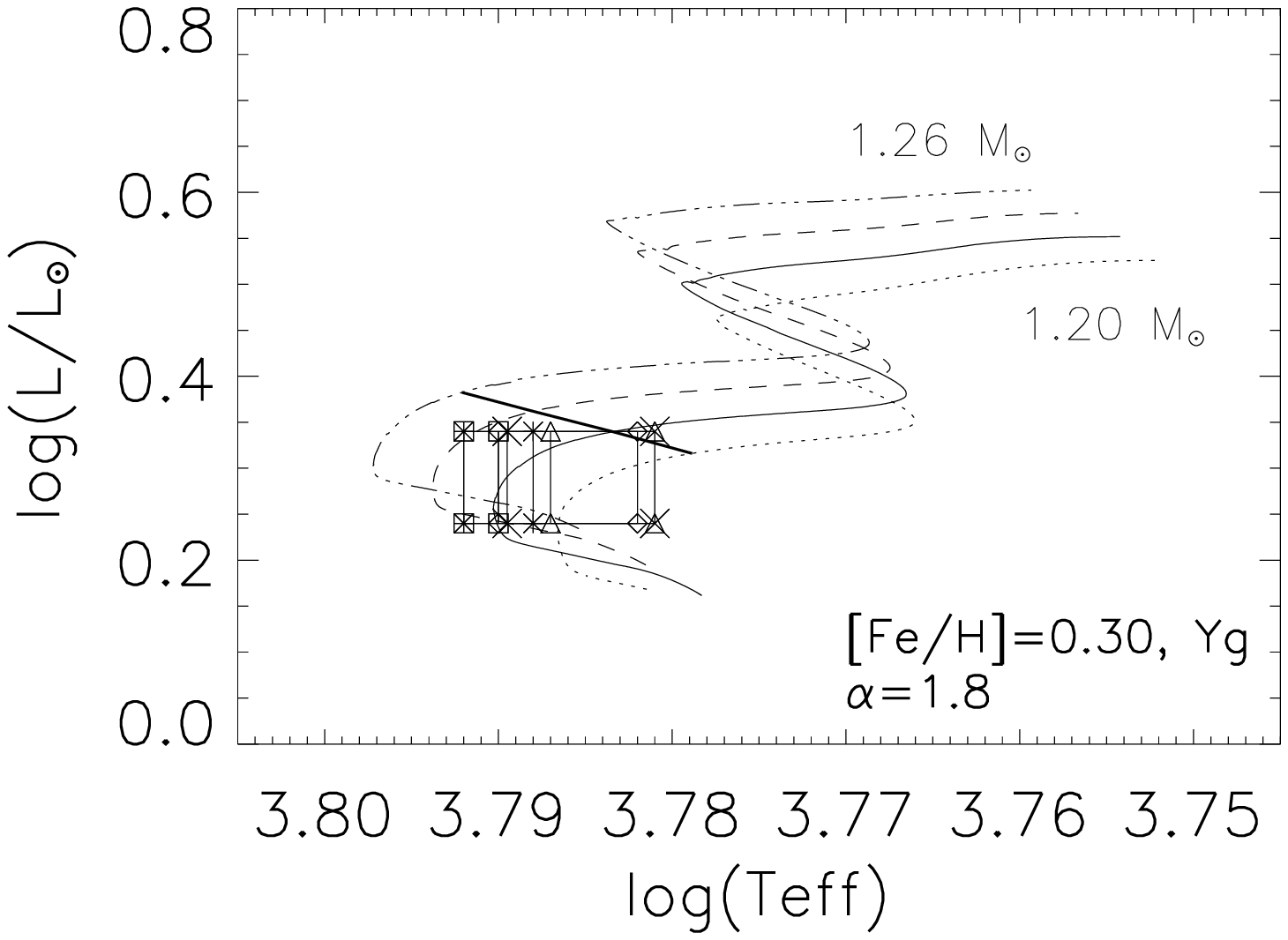}
   \includegraphics[width=0.48\textwidth]{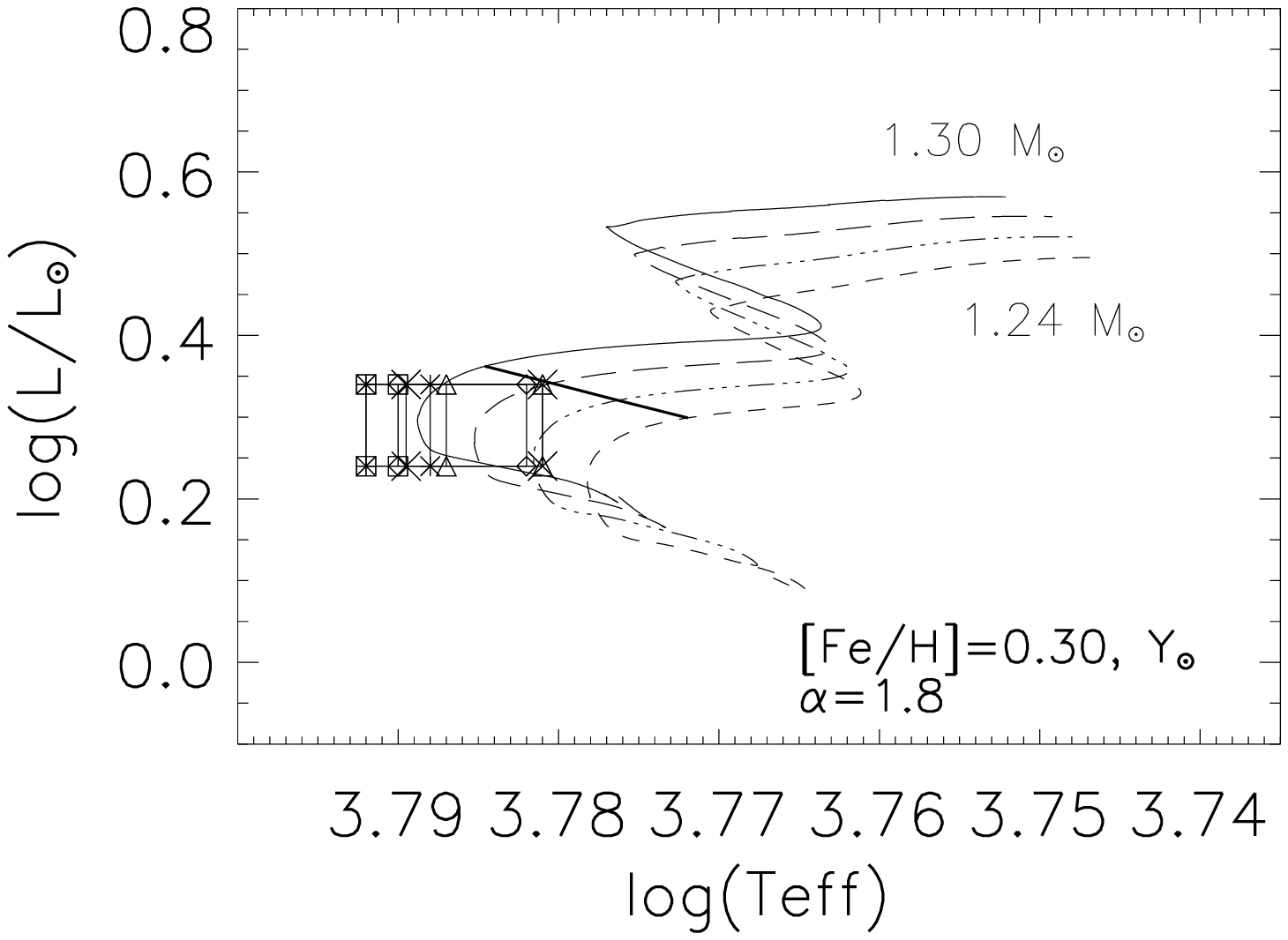}
     \caption{HR diagrams for $\alpha=1.8$ (see text for details). Error boxes are the same horizontally as presented in Figure~\ref{logtracks1}, for the effective temperatures, and correspond vertically to the luminosity uncertainty.
     The straight thick line represents the iso-$<\Delta \nu>$ line, with $<\Delta \nu>$=98.4 $\mu$Hz.}
              \label{logtracks2}
    \end{figure*}

%%%%%%%%%%%%%%%%%%%%%%%%%%%%%%%%%%%%%%%%%%%%%%%%%%%%%%%%%%%%%%%%%%%%%%%%%%%%%%%%%%%%%%%%%%%%%%%%%%%%%%%%%%%%%%%%%%%%%%%%%%%%%%

\subsection{Test on atomic diffusion: radiative accelerations}

Atomic diffusion is a very important process inside stars: it can modify the atmospheric abundances and also
have strong implications for the internal structure of stars (Richard, Michaud \& Richer \cite{richard01}; 
Michaud et al. \cite{michaud04}; Th\'eado et al. \cite{theado09}). 
At the present time, most stellar evolution codes include the computation of gravitational settling, but not the 
computation of radiative accelerations of heavy elements. These accelerations, which oppose gravitation, are negligible 
for the Sun but they rapidly become important for more massive stars (Michaud et al. \cite{michaud76}). 
The variations with depth of the radiative accelerations on specific elements can lead to their accumulation or depletion in various
layers inside stars. The presence of a heavy iron layer above layers with smaller molecular weights creates an inverse $\mu$-gradient,
unstable towards thermohaline convection (Vauclair \cite{vauclair04}). This induced mixing has also to be taken into account
in computations of stellar modeling (Th\'eado et al. \cite {theado09}).
An improved TGEC version including radiative accelerations
on C, N, O, Ca and Fe was recently developped (Th\'eado et al. \cite{theado12}). This new version was used to compare 
the oscillation frequencies computed with and without introducing the radiative accelerations in the models. We found 
that, for solar-type stars (masses less than 1.30 M$_{\sun}$), the difference in the computed frequencies is small. Two models 
with the same mass of 1.28  M$_{\sun}$ and the same average large separations of 98.26 $\mu$Hz, one computed with the 
radiative accelerations and one neglecting them, present differences in the average small separations of the order of 0.01 $\mu$Hz. These 
differences between the two models decrease with decreasing stellar mass. We conclude that the radiative accelerations may be neglected 
in the following computations.

The result that radiative accelerations have no important consequences in the present models is consistent with the fact that the 
detailed abundances do not show large relative variations for the observed elements (Section 1). Indeed, in case of gravitational 
settling, the abundances of heavy elements all decrease in a similar way, in spite of their different masses, due to the slowing 
down effect of the coulomb forces. The diffusion velocities vary typically as A/Z$^2$ where A is the mass number and Z the charge, 
which is similar for various elements in stellar conditions. This behavior would be modified if radiative accelerations were important.

%%%%%%%%%%%%%%%%%%%%%%%%%%%%%%%%%%%%%%%%%%%%%% TABLA 2 + Teff and L/Lsun %%%%%%%%%%%%%%%%%%%%%%%%%%%%%%%%%%%%%%%%%%%%%%%

\begin{table*}
\caption{Examples of models with $\alpha$=1.8, without surface effects.}
\label{fe019yg}
\centering
\begin{tabular}{c c c c c c c c c c c c c}     % 12 columns
\hline\hline
[Fe/H]$_{i}$ & Y$_{i}$ & M/M$_{\sun}$ & Age & [Fe/H]$_{S}$ & Y$_{S}$ & log $g$ & log $T_\mathrm{eff}$ & log $(L/L_{\sun})$ & R/R$_{\sun}$ & 
M/R$^{3}$ & $<\Delta \nu>$  & $<\delta \nu_{02}>$ \\ %& $\chi^{2}$ \\
  &  &  & [Gyr] &  &  & [K] &  &  &  & [solar units] & [$\mu$Hz] & [$\mu$Hz] \\
\hline

0.23 & 0.293 & 1.18 & 3.682 & 0.16 & 0.246 & 4.267 & 3.778 & 0.310 & 1.328 & 0.50 & 98.31 & 7.08 \\ 
0.23 & 0.293 & 1.20 & 3.204 & 0.16 & 0.246 & 4.271 & 3.782 & 0.332 & 1.333 & 0.51 & 98.38 & 7.54 \\ 
0.23 & 0.293 & 1.22 & 2.820 & 0.16 & 0.248 & 4.273 & 3.787 & 0.355 & 1.341 & 0.51 & 98.36 & 7.85 \\ 
0.23 & 0.293 & 1.24 & 2.416 & 0.15 & 0.242 & 4.276 & 3.791 & 0.375 & 1.347 & 0.51 & 98.34 & 8.33 \\ 

\hline
\hline

0.23 & 0.271 & 1.22 & 3.756 & 0.16 & 0.228 & 4.272 & 3.776 & 0.312 & 1.343 & 0.51 & 98.33 & 7.11 \\
0.23 & 0.271 & 1.24 & 3.283 & 0.16 & 0.228 & 4.275 & 3.780 & 0.334 & 1.349 & 0.51 & 98.31 & 7.80 \\
0.23 & 0.271 & 1.26 & 2.865 & 0.16 & 0.230 & 4.278 & 3.785 & 0.355 & 1.355 & 0.51 & 98.34 & 7.98 \\
0.23 & 0.271 & 1.28 & 2.461 & 0.16 & 0.227 & 4.281 & 3.789 & 0.375 & 1.361 & 0.51 & 98.35 & 8.30 \\

\hline
\hline

0.30 & 0.303 & 1.20 & 3.209 & 0.23 & 0.257 & 4.271 & 3.778 & 0.316 & 1.333 & 0.51 & 98.42 & 7.35 \\
0.30 & 0.303 & 1.22 & 2.820 & 0.23 & 0.258 & 4.273 & 3.783 & 0.339 & 1.341 & 0.51 & 98.35 & 7.75 \\
0.30 & 0.303 & 1.24 & 2.431 & 0.23 & 0.261 & 4.276 & 3.787 & 0.361 & 1.347 & 0.51 & 98.38 & 8.11 \\
0.30 & 0.303 & 1.26 & 2.072 & 0.22 & 0.254 & 4.282 & 3.792 & 0.382 & 1.354 & 0.51 & 98.38 & 8.59 \\
%0.30 & 0.3027 & 1.28 & 1.673 & 0.22 & 0.256 & 4.285 & 3.795 & 0.399 & 1.358 & 0.51 & 98.33 & 8.87 \\

\hline
\hline

%0.30 & 0.2714 & 1.20 & 4.878 & 0.23 & 0.231 & 4.270 & 3.763 & 0.255 & 1.335 & 0.51 & 98.42 & 6.17 \\
%0.30 & 0.2714 & 1.22 & 4.280 & 0.23 & 0.231 & 4.272 & 3.767 & 0.276 & 1.341 & 0.51 & 98.42 & 6.68 \\
0.30 & 0.271 & 1.24 & 3.771 & 0.23 & 0.231 & 4.274 & 3.771 & 0.299 & 1.349 & 0.51 & 98.32 & 7.11 \\
0.30 & 0.271 & 1.26 & 3.293 & 0.23 & 0.231 & 4.277 & 3.776 & 0.320 & 1.356 & 0.51 & 98.37 & 7.64 \\
0.30 & 0.271 & 1.28 & 2.865 & 0.23 & 0.231 & 4.280 & 3.780 & 0.341 & 1.363 & 0.51 & 98.34 & 7.87 \\
0.30 & 0.271 & 1.30 & 2.476 & 0.24 & 0.233 & 4.282 & 3.784 & 0.362 & 1.369 & 0.51 & 98.39 & 8.25 \\

\hline
\end{tabular}
\end{table*}
%%%%%%%%%%%%%%%%%%%%%%%%%%%%%%%%%%%%%%%%%%%%%%%%%%%%%%%%%%%%%%%%%%%%%%%%%%%

\section{Results}

\subsection{Computations without surface effects }

The computations of evolutionary tracks, with two initial metallicity values, [Fe/H] = 0.23 and 0.30, and two different helium abundances, Y$_{\sun}$ equal to 0.271 and Y$_{G}$ respectively equal to 0.293 and 0.303 for the two metallicity values, lead to four different sets of tracks, each of them covering a mass range from 1.10 to 1.30 $M_{\sun}$. A few of these tracks are presented in Figures 1 and 2. Error boxes of five of the spectroscopic studies given in Table 1 are also drawn in these figures.

As explained in previous sections, we found, along each evolutionary track, a model which has an average large separation consistent with the observed one,
computed in the same frequency range, of $\sim$ [1.5, 2.5] mHz.
The location of all these models in the log g - log $T_\mathrm{eff}$ plane as well as in the HR diagrams
are indicated in Figures~\ref{logtracks1} and \ref{logtracks2} with iso-$<\Delta \nu>$ 98.4 $\mu$Hz lines. For each case, we also computed the average small separation,  $\delta \nu _{n,02}$ (Table 2). We can see that for models with the same large separation, the small separation increases for increasing mass, so that in each case (i.e. for each set of chemical composition), there is a model that is consistent with both the large and small separations. However, when comparing the absolute model frequencies with the observed ones, we find that we must shift the computed frequencies by about 20 $\mu$Hz to obtain the best fit with the observed ones. This offset is attributed to surface effects.  The \'echelle diagrams corresponding to these best models are given in Figure 3.

%%%%%%%%%%%%%%%%%%%%%%%%%%%%%%% FIGURE EDs ALPHA 1.8 %%%%%%%%%%%%%%%%%%%%%%%%%%%%%%%%%%%%%%%%%%%%%%%%%%%%%%%%%

   \begin{figure*}
   \centering
   \includegraphics[width=0.5\textwidth]{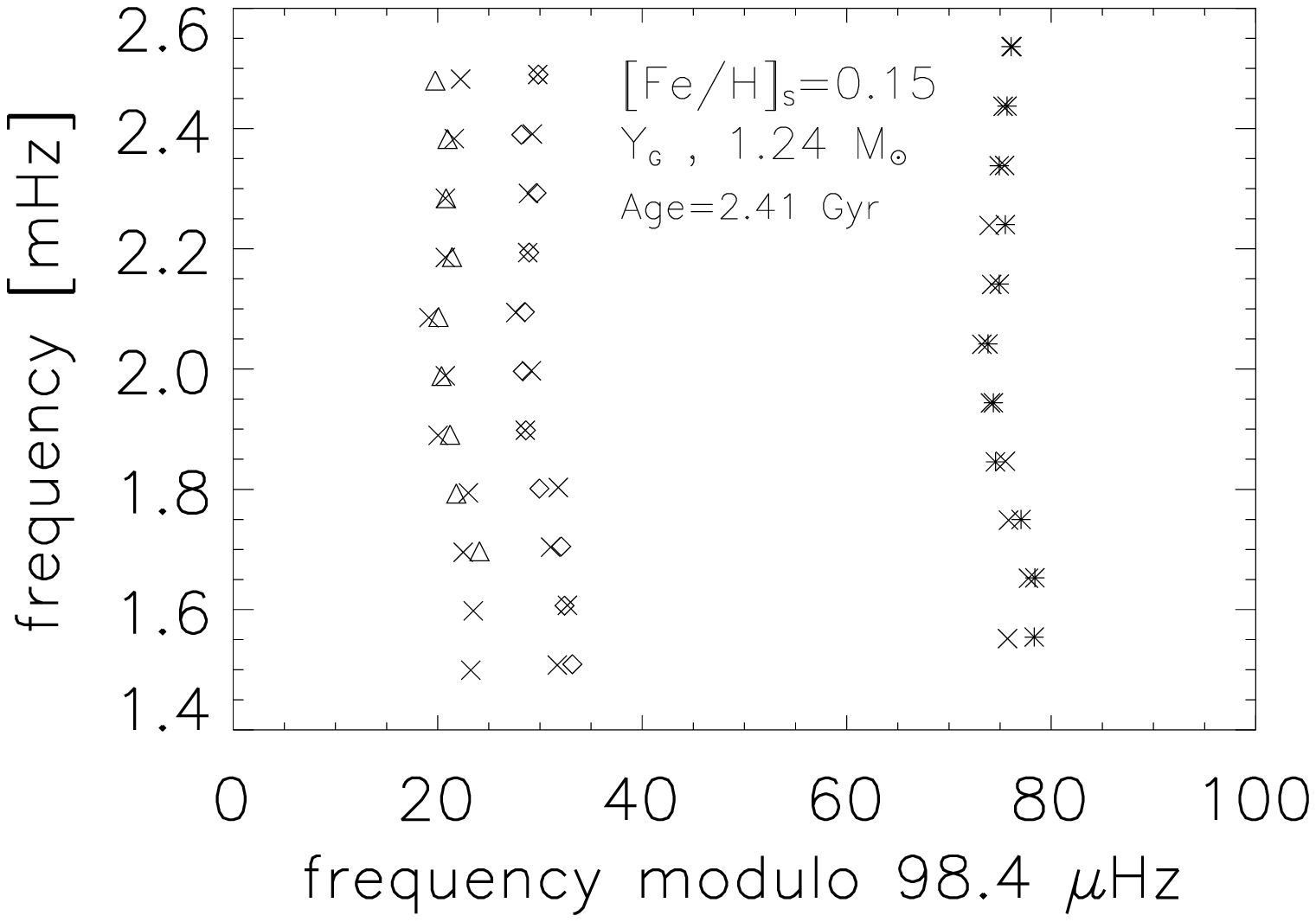}%
   \includegraphics[width=0.5\textwidth]{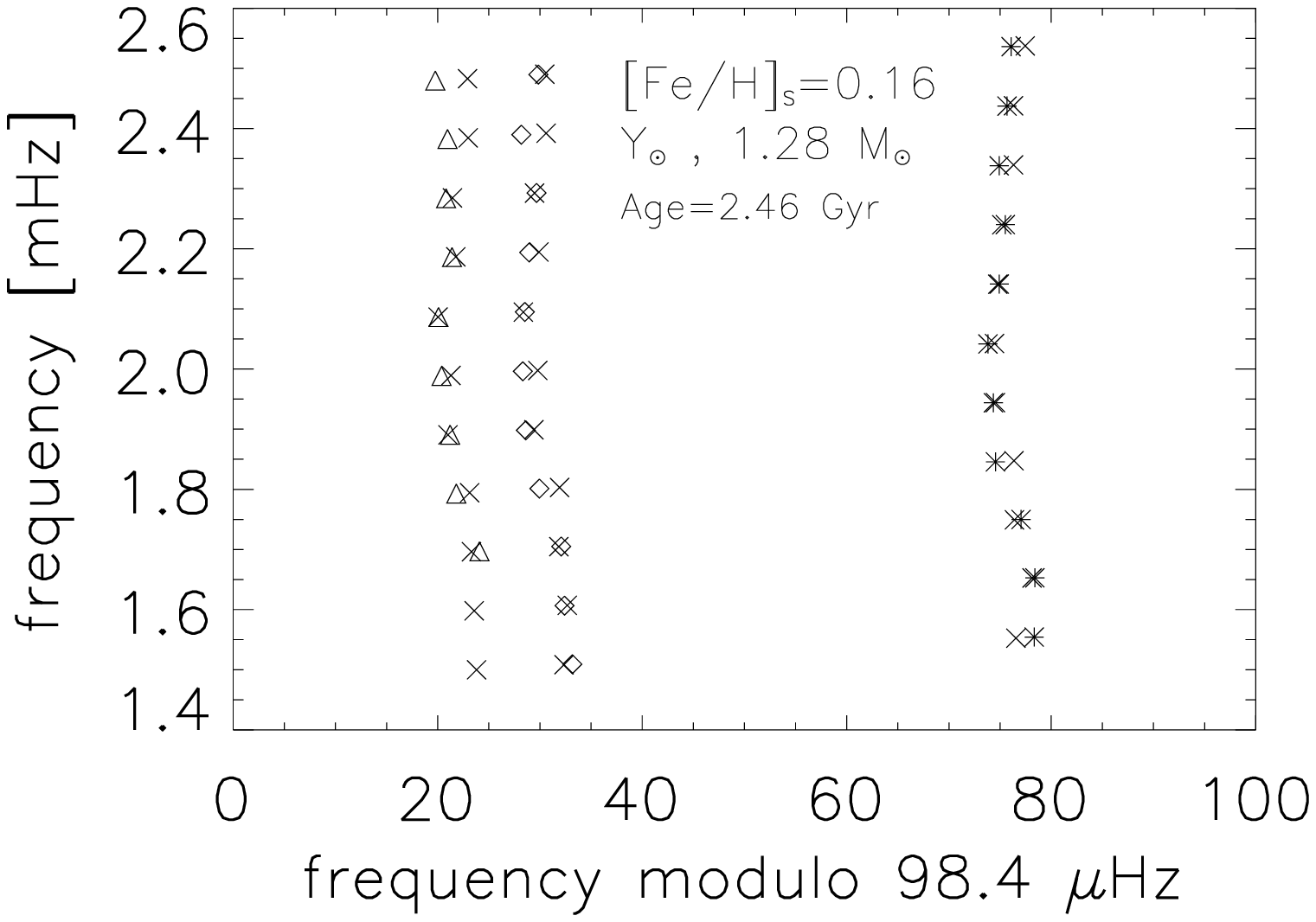}
   \includegraphics[width=0.5\textwidth]{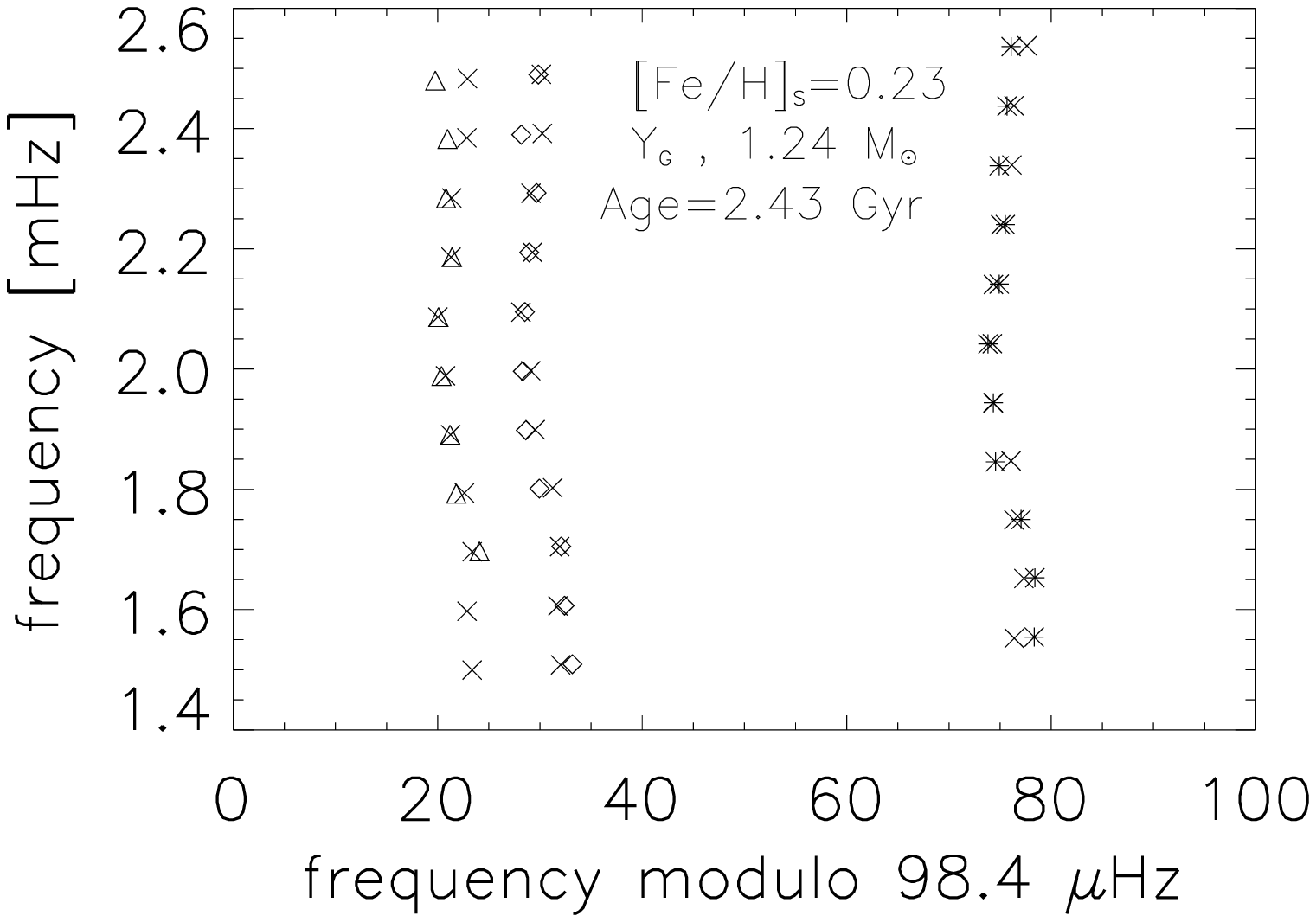}%
   \includegraphics[width=0.5\textwidth]{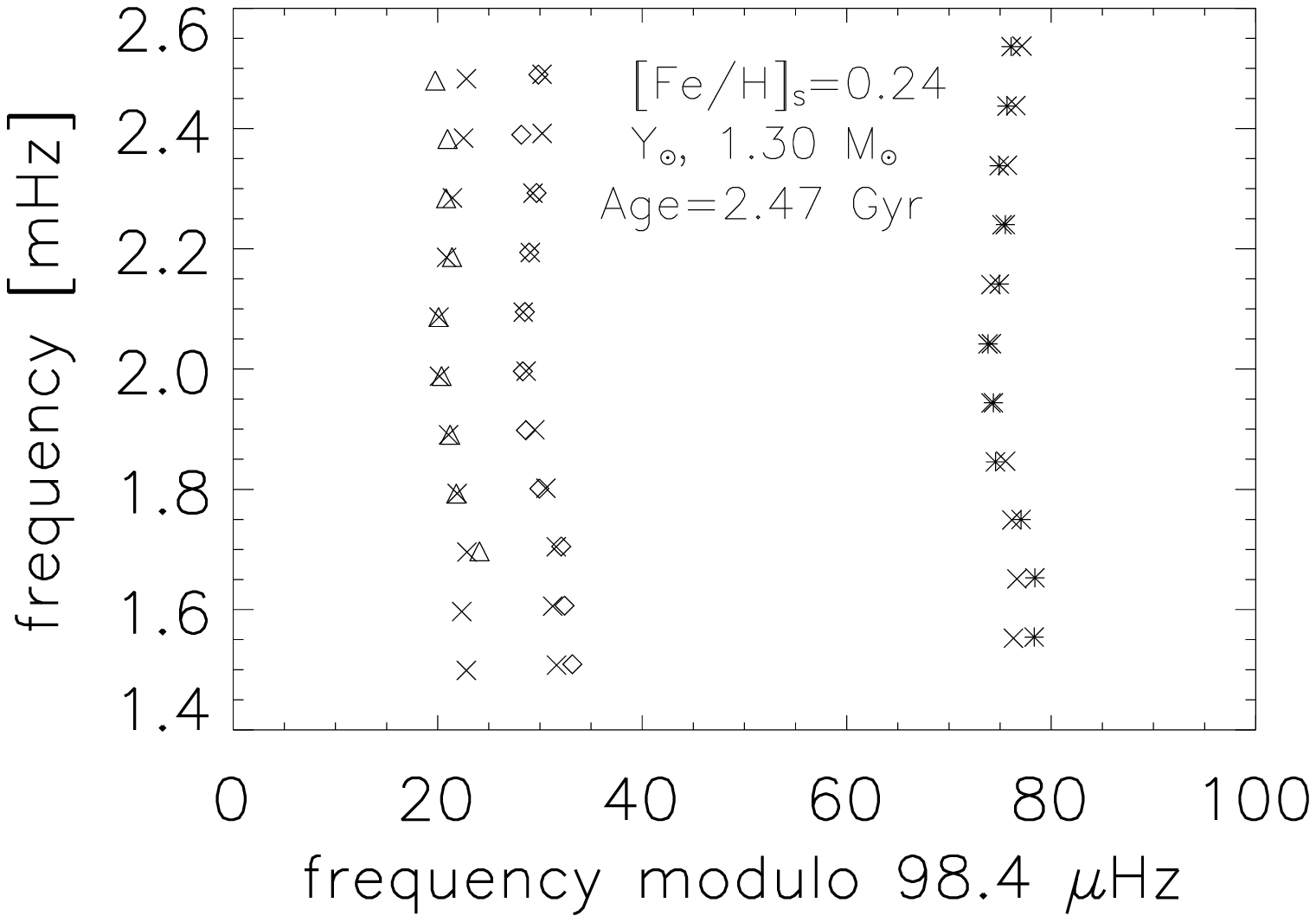}
     \caption{Echelle diagrams for the best model found for each set of evolutionary tracks calculated with $\alpha=1.8$. In this set of models, the surface effects are not included. The 
model frequencies (\textit{crosses}) are compared to the observed frequencies, represented as diamonds ($\ell$=0), asterisks ($\ell$=1), and triangles ($\ell$=2).To obtain these best fits we had to shift the model frequencies by respectively (top left, top right, bottom left, bottom right) 27, 25, 24, 22 $\mu$Hz (see text)}
   \label{edfitm181}
    \end{figure*}

%%%%%%%%%%%%%%%%%%%%%%%%%%%%%%%%%%%%%%%%%%%%%%%%%%%%%%%%%%%%%%%%%%%%%%%%%%%%%%%%%%%%%%%%%%%%%%%%%%%%%%%%%%%%%%%%%%%%%%%%%%%%%%

%%%%%%%%%%%%%%%%%%%%%%%%%%%%%%%%%%%%%%%%%%%%%% TABLE FE019YG SURFACE EFFECTS %%%%%%%%%%%%%%%%%%%%%%%%%%%%%%%%%%%%%%%%%%%%%%%

\begin{table*}
\caption{Examples of models with  $\alpha$=1.8 , including surface effects. Here $<\Delta \nu^*>$ represents the Kjeldsen et al. (\cite{kjeldsen08})-corrected large separations.}
\label{fe019ygse}
\centering
\begin{tabular}{c c c c c c c c c c c c c c}     % 12 columns
\hline\hline
[Fe/H]$_{i}$ & Y$_{i}$ & M/M$_{\sun}$ & Age & [Fe/H]$_{S}$ & Y$_{S}$ & log $g$ & log $T_\mathrm{eff}$ & log $(L/L_{\sun})$ & R/R$_{\sun}$ &
M/R$^{3}$ & $<\Delta \nu^*>$ & $<\Delta \nu>$ & $<\delta \nu_{02}>$ \\ % & $\chi^{2}$ \\
  &  &  & [Gyr] &  &  & [K] &  &  &  & [solar units] & [$\mu$Hz] & [$\mu$Hz] & [$\mu$Hz] \\
\hline
0.23 & 0.293 & 1.18 & 3.548 & 0.16 & 0.247 & 4.275 & 3.779 & 0.306 & 1.315 & 0.52 & 99.67 & 98.27 & 7.27 \\ 
0.23 & 0.293 & 1.20 & 3.069 & 0.16 & 0.247 & 4.279 & 3.784 & 0.328 & 1.321 & 0.52 & 99.68 & 98.27 & 7.83 \\ 
0.23 & 0.293 & 1.22 & 2.670 & 0.16 & 0.250 & 4.282 & 3.788 & 0.350 & 1.327 & 0.52 & 99.86 & 98.19 & 8.11 \\ 
0.23 & 0.293 & 1.24 & 2.251 & 0.16 & 0.244 & 4.287 & 3.792 & 0.369 & 1.331 & 0.52 & 100.08& 98.19 & 8.47 \\ 

\hline
\hline

0.23 & 0.271 & 1.20 & 4.130 & 0.16 & 0.228 & 4.276 & 3.773 & 0.287 & 1.325 & 0.52 & 99.51 & 98.19 & 6.85 \\
0.23 & 0.271 & 1.22 & 3.622 & 0.16 & 0.228 & 4.279 & 3.777 & 0.309 & 1.331 & 0.52 & 99.56 & 98.18 & 7.26 \\ 
0.23 & 0.271 & 1.24 & 3.149 & 0.16 & 0.229 & 4.283 & 3.781 & 0.330 & 1.337 & 0.52 & 99.61 & 98.18 & 7.69 \\ 
0.23 & 0.271 & 1.26 & 2.730 & 0.16 & 0.232 & 4.286 & 3.786 & 0.351 & 1.343 & 0.52 & 99.62 & 98.07 & 8.11 \\ 

\hline
\hline

0.30 & 0.303 & 1.20 & 3.089 & 0.23 & 0.258 & 4.278 & 3.780 & 0.313 & 1.322 & 0.52 & 99.64 & 98.32 & 7.70 \\
0.30 & 0.303 & 1.22 & 2.685 & 0.23 & 0.260 & 4.282 & 3.784 & 0.335 & 1.328 & 0.52 & 99.72 & 98.13 & 7.94 \\
0.30 & 0.303 & 1.24 & 2.296 & 0.24 & 0.263 & 4.285 & 3.788 & 0.357 & 1.334 & 0.52 & 99.78 & 98.15 & 8.34 \\
0.30 & 0.303 & 1.26 & 1.907 & 0.23 & 0.257 & 4.290 & 3.793 & 0.376 & 1.337 & 0.52 & 100.23& 98.20 & 8.84 \\

\hline
\hline

0.30 & 0.271 & 1.24 & 3.637 & 0.23 & 0.231  &4.282 & 3.773 & 0.295 & 1.338 & 0.52 & 99.56 & 98.27 & 7.30 \\
0.30 & 0.271 & 1.26 & 3.173 & 0.23 & 0.232 & 4.284 & 3.777 & 0.317 & 1.345 & 0.52 & 99.50 & 98.16 & 7.64 \\
0.30 & 0.271 & 1.28 & 2.730 & 0.23 & 0.233 & 4.288 & 3.781 & 0.337 & 1.351 & 0.52 & 99.64 & 98.09 & 8.08 \\
0.30 & 0.271 & 1.30 & 2.326 & 0.23 & 0.236 & 4.291 & 3.785 & 0.357 & 1.355 & 0.52 & 99.80 & 98.20 & 8.52 \\

\hline
\end{tabular}
\end{table*}

%%%%%%%%%%%%%%%%%%%%%%%%%%%%%%%%%%%%%%%%%%%%%%%%%%%%%%%%%%%%%%%%%%%%%%%%%%%%%%%

\subsection{Computations including surface effects}

\begin{figure*}
   \centering
   \includegraphics[width=0.5\textwidth]{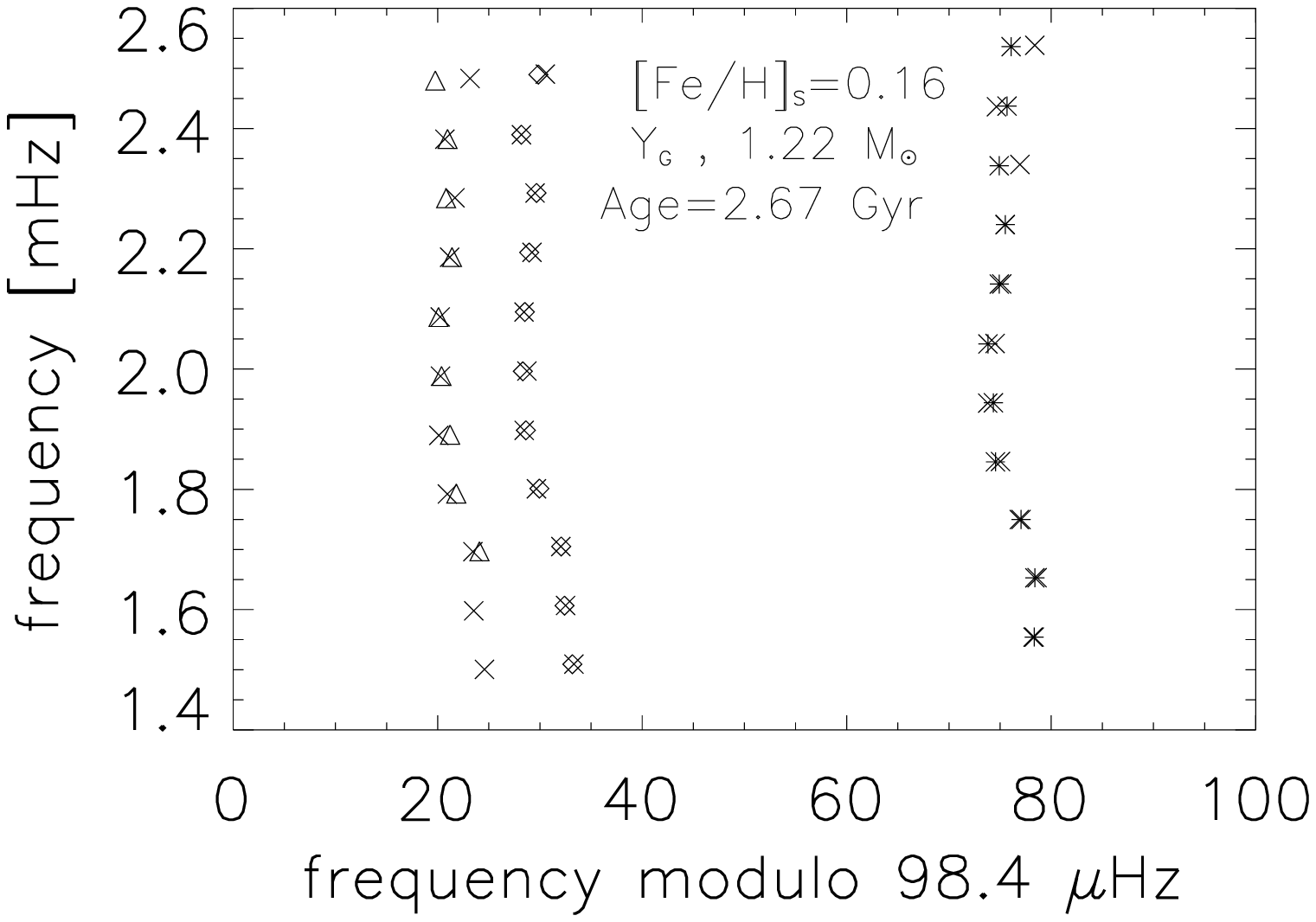}%
   \includegraphics[width=0.5\textwidth]{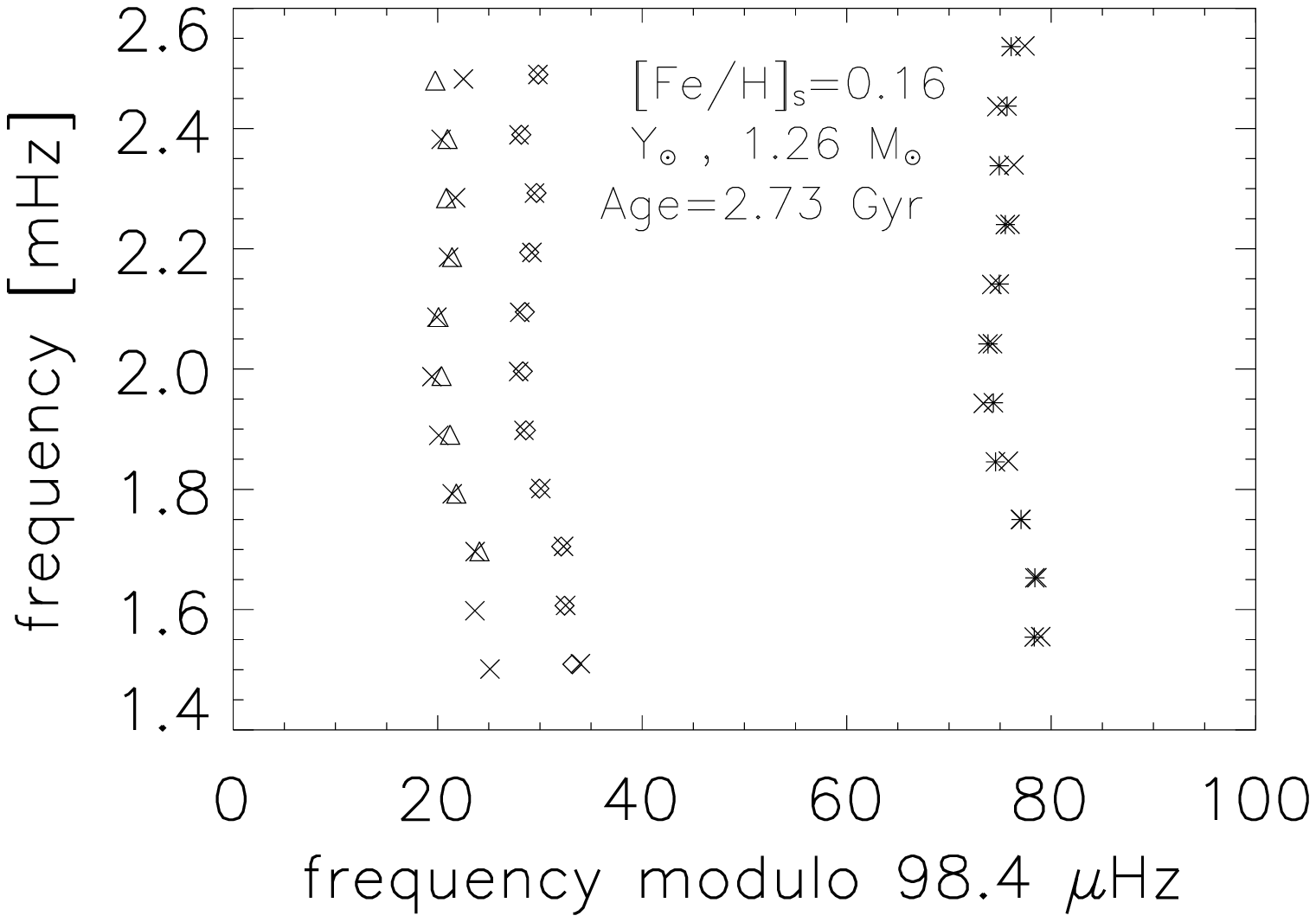}
   \includegraphics[width=0.5\textwidth]{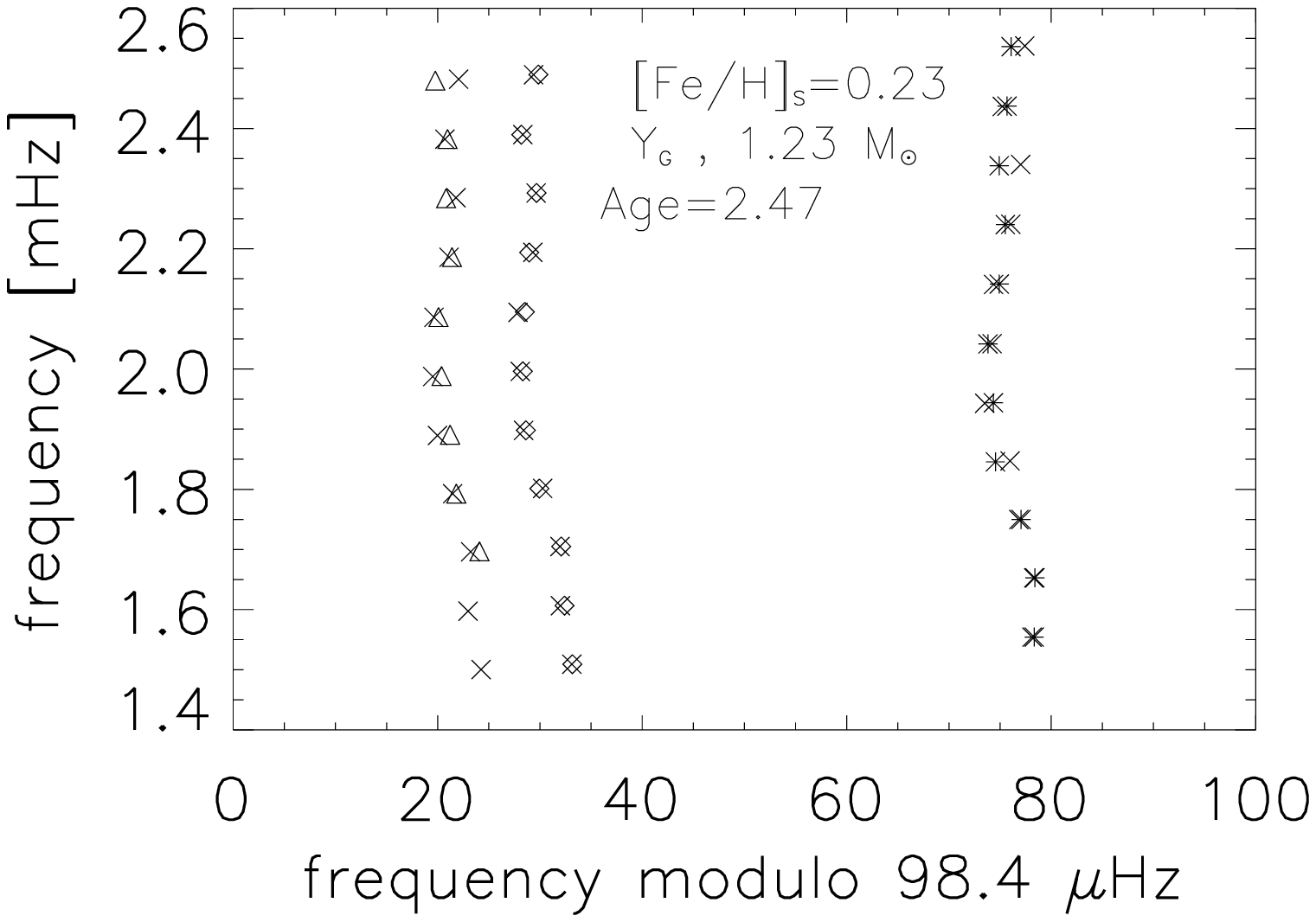}%
   \includegraphics[width=0.5\textwidth]{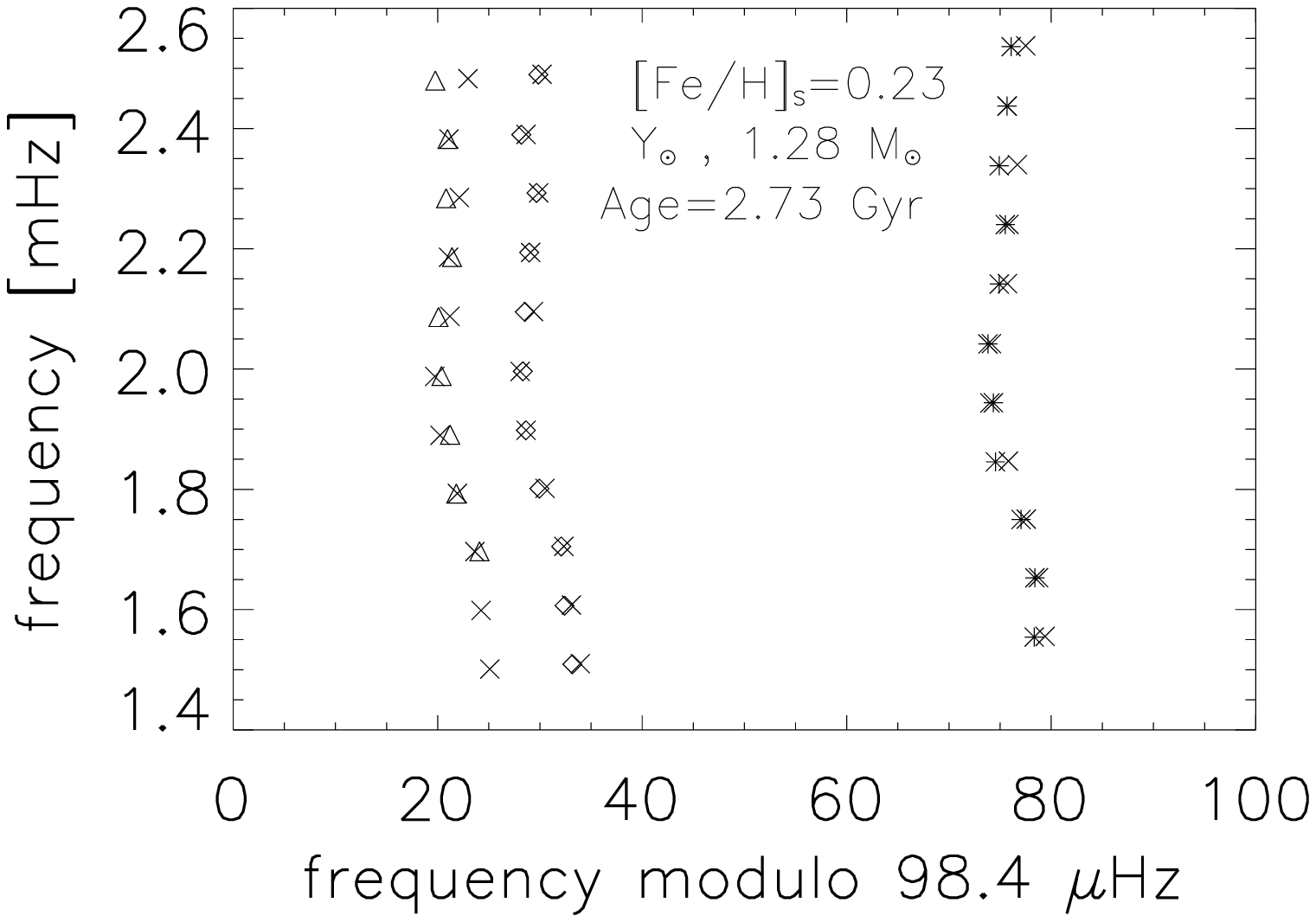}
     \caption{Echelle diagrams for the best model, including near-surface corrections, as proposed by Kjeldsen et al. \cite{kjeldsen08}
 found for each set of evolutionary tracks calculated with $\alpha=1.8$, in comparison with
observed frequencies. The symbols are the same ones as in Figure 3.}
   \label{edse18}
    \end{figure*}

We know that stellar modeling fails to represent properly the near-surface layers of the stars. As a consequence,
there is a systematic offset between observed and computed frequencies. This offset is independent of the angular degree $\ell$ and
increases with frequency. This has been studied in the case of the Sun, and a similar offset is expected to occur
in other stars. Using the Sun as a reference, Kjeldsen et al. (\cite{kjeldsen08}) suggested that
for other stars, the near-surface correction on the frequencies may be approximated by:

\begin{equation}
    \nu _\mathrm{obs}(n) - \nu_\mathrm{best}(n) = a \left[\frac{\nu_\mathrm{obs}(n)}{\nu_{0}}\right]^{b}
\end{equation}
where $\nu_\mathrm{obs}(n)$ are the observed $\ell$=0 frequencies with radial order $n$, $\nu_\mathrm{best}(n)$ are the calculated frequencies for
the best model which is the model that best describes the star but which still fails to model correctly
the near-surface layers, and $\nu_{0}$ is a constant reference frequency, chosen to be that of the frequency at the maximum amplitude in the power spectrum. The parameter $a$ may be derived as a function of the parameter $b$, which has to be adjusted to the solar case.
We used this method to find the frequency corrections that we have to apply to our models, which lead to a new average large separation $<\Delta \nu^*>$, slightly larger than the observed one.
We then proceed as before to derive models for the same sets of values of the helium abundance and metallicity, using $\alpha = 1.8$ (Table 3).
Figure~\ref{edse18} presents \'echelle diagrams obtained with the new best models. In these graphs, the frequency corrections have been applied to the new model frequencies in order to compare them directly with the observations.

%%%%%%%%%%%%%%%%%%%%%%%%%%%%%%% FIGURE EVOLUTIONARY TRACKS ALPHA 1.8 %%%%%%%%%%%%%%%%%%%%%%%%%%%%%%%%%%%%%%%%%%%%%%%%%%%%%%%%%

%%%%%%%%%%%%%%%%%%%%%%%%%%%%%%%%%%%%%%%%%%%%%%%%%%%%%%%%%%%%%%%%%%%%%%%%%%%%%%%%%%%%%%%%%%%%%%%%%%%%%%%%%%%%%%%%%%%%%%%%%%%%%%

\subsection{Best models and discussion}

We give in Table~\ref{fe023yg} the parameters of the best models that we obtained for the four different sets of chemical composition. These models have been computed with a mixing length parameter of 1.8, the frequencies have been corrected for surface effects and the computed large and small separations are the closest to the observed ones in the sample. We also give the $\chi^2$ values for the comparisons of the three $\ell$ = 0, 1, 2 lines in the \'echelle diagrams. 

\begin{figure}[h!]
   \centering
   \includegraphics[width=0.5\textwidth]{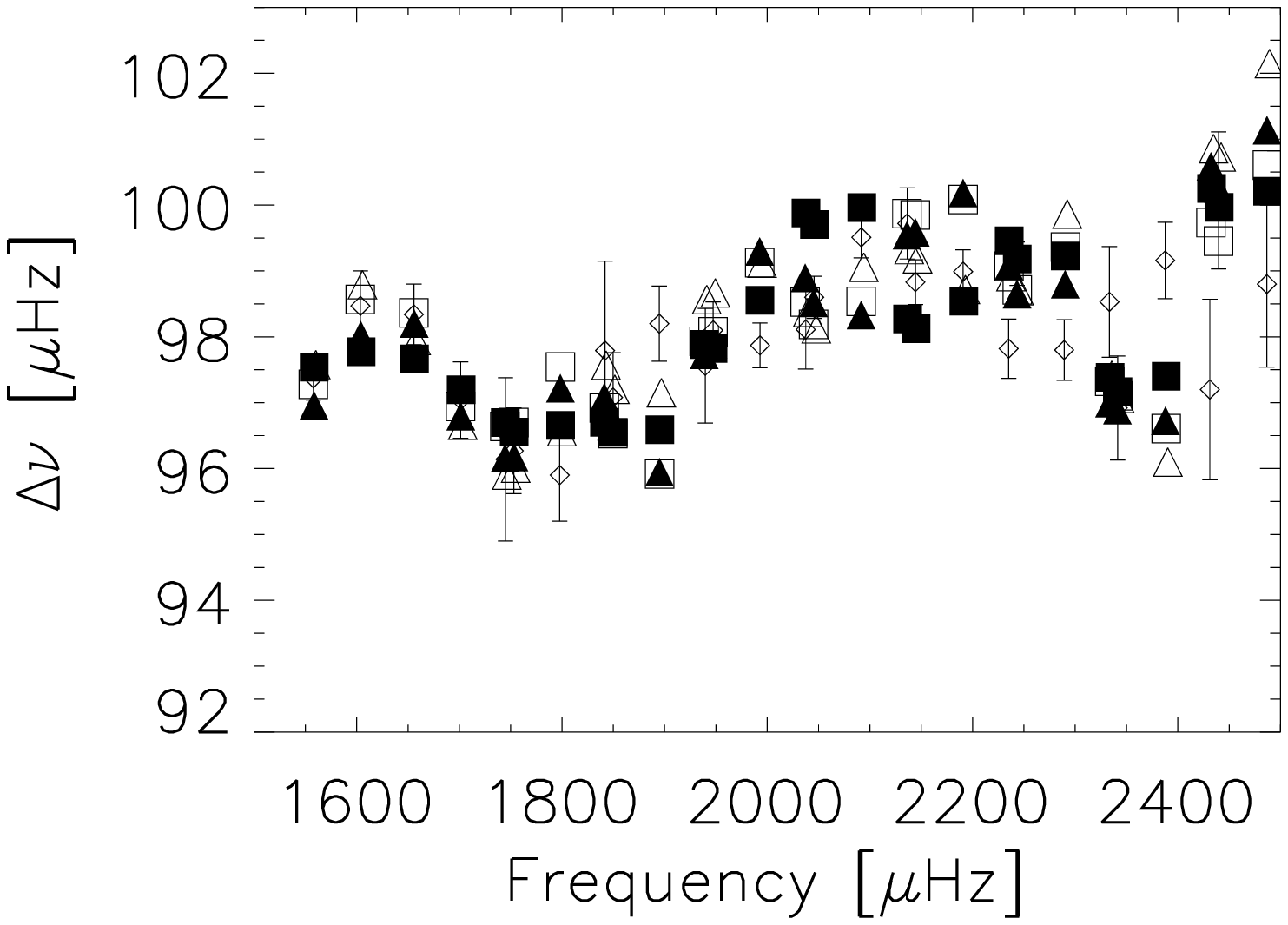}
   \includegraphics[width=0.5\textwidth]{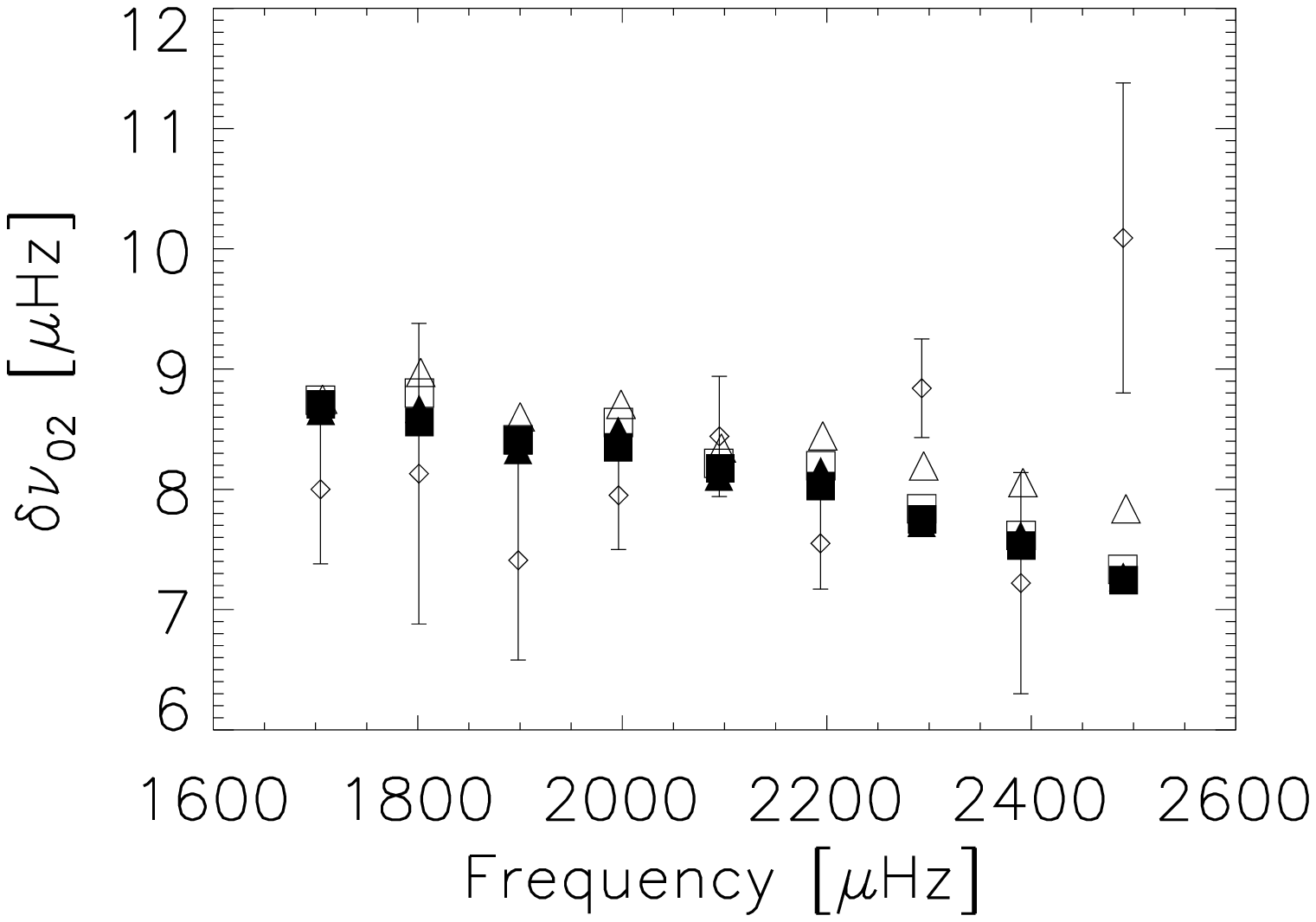}
       \caption{Comparisons between the large separations (top) and the small separations (bottom) of the four best models 
indicated by squares for models with [Fe/H]$_{s}$=0.23 ([Fe/H]$_{i}$=0.30) and triangles for models with 
[Fe/H]$_{s}$=0.16 ([Fe/H]$_{i}$=0.23). Empty symbols are used to indicated an 
helium abundance according to Y$_{g}$,and filled symbols for a solar helium abundance (see text for details.) Observations are represented as white diamonds.}
%       \caption{Comparisons between the large separations (left) and the small separations (right) of the four best models (black and white squares and triangles, see Figure 6 for details) and the observations (white diamonds).}
              \label{comp-best}
    \end{figure}

Figure~\ref{comp-best} displays the large and small separations as a function of the frequency for the observations and the four best models. The pattern observed in the large separations are well reproduced by the models. For the small separations, the agreement is also very good except for two points at large frequencies (2284 $\mu$Hz and 2479 $\mu$Hz). This suggests that the uncertainties given in Ballot et al. (\cite{ballot11}) for these points were underestimated. 

Several concluding points can already be derived from a first analysis of Table 4. First of all, the stellar gravity is obtained, as usual, with a precision of order $0.1\%$.  The mass and age depend basically on the chosen value for the initial helium content. 
For a low helium value, the mass is 1.26 to 1.28 $M_{\sun}$, and the age 2.73 Gyr,
whereas  for a larger helium value the mass is slightly smaller (around 1.22 - 1.23 $M_{\sun}$) as well as the age (2.48 to 2.68 Gyr).
In any case the radius and luminosity are known with a precision of order $1\%$. 

We can go further by comparing the effective temperatures of the models with the spectroscopic observations (Figure 6). 
As found before for the cases of $\iota$ Hor (Vauclair et al. \cite{vauclair08}) and $\mu$ Arae (Soriano \& Vauclair \cite{soriano10}), 
the effective temperatures of the best models are smaller for smaller initial helium and larger for smaller metallicity. 
In the present case, we find that the model with the largest metallicity and the smallest helium, represented by black square in figure 6, is at the coolest 
limit of the observational boxes and thus may be excluded from the sample on spectroscopic considerations. 
We derive the stellar parameters from a mean value of the resulting three models, as given in Table 5.

%%%%%%%%%%%%%%%%%%%%%%%%%%%%%%%%%%%%%%%%%%% FINAL BEST MODELS SURFACE EFFECTS %%%%%%%%%%%%%%%%%%%%%%%%%%%%%%%%%%%%%%%%%%%%%%%%

\begin{table*}
\caption{Best models obtained with the TGEC code, including near surface corrections.}
\label{fe023yg}
\centering
\begin{tabular}{c c c c c c c c c c c }     % 9 columns
\hline\hline
M/M$_{\sun}$ & $L/L_{\sun}$ & R/R$_{\sun}$ & log $g$ & T$_\mathrm{eff}$ [K] & age [Gyr] & [Fe/H]$_{i}$ & Y$_{i}$ & [Fe/H]$_{S}$ & Y$_{S}$ & $\chi^{2}$ \\
\hline

1.22 & 2.239 & 1.327 & 4.282 & 6143 & 2.670 & 0.23 & 0.293 & 0.16 & 0.250 & 5.06 \\
1.23 & 2.219 & 1.330 & 4.283 & 6120 & 2.476 & 0.30 & 0.303 & 0.23 & 0.262 & 3.34 \\
1.26 & 2.244 & 1.343 & 4.286 & 6109 & 2.730 & 0.23 & 0.271 & 0.16 & 0.232 & 6.42 \\
1.28 & 2.173 & 1.351 & 4.288 & 6043 & 2.730 & 0.30 & 0.271 & 0.23 & 0.233 & 3.51 \\

\hline
\end{tabular}
\end{table*}

%%%%%%%%%%%%%%%%%%%%%%%%%%%%%%%%%%%%%%%%%%%%%%%%%%%%%%%%%%%%%%%%%%%%%%%%%%%%%%%%%%%%%%%%%%%%%%%%%%%%%%%%%%%%%%%%
%%%%%%%%%%%%%%%%%%%%%%%%%%%%%%% FIGURE COMPARISON MODELS %%%%%%%%%%%%%%%%%%%%%%%%%%%%%%%%%%%%%%%%%%%%%%%%%%%%%%%%%

%%%%%%%%%%%%%%%%%%%%%%%%%%%%%%%%%%%%%%%%%%%%%%%%%%%%%%%%%%%%%%%%%%%%%%%%%%%%%%%%%%%%%%%%%%%%%%%%%%%%%%%%%%%%%%%%%%%%%%%%%%%%%%

%%%%%%%%%%%%%%%%%%%%%%%%%%%%%%% FIGURE ERROR BOXES - BEST MODELS ALPHA 1.8 SE %%%%%%%%%%%%%%%%%%%%%%%%%%%%%%%%%%%%%%%%%%%%%%%%%%%%

   \begin{figure}
   \centering
   \includegraphics[width=0.5\textwidth]{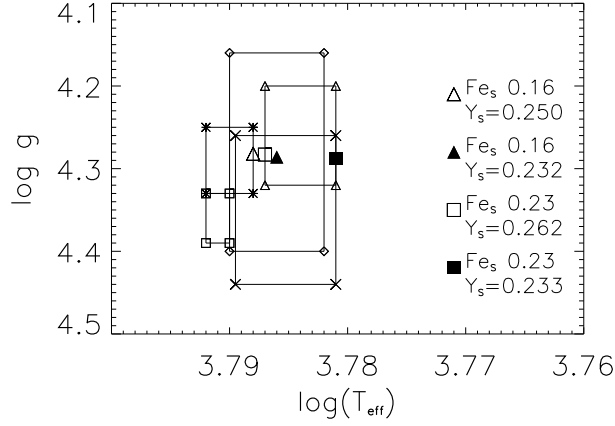}
     \caption{Location of the best models in the log g vs log $T_\mathrm{eff}$ plane. Triangles indicate models with [Fe/H]$_{s}$=0.16 ([Fe/H]$_{i}$=0.23), 
and squares indicate  models with [Fe/H]$_{s}$=0.23 ([Fe/H]$_{i}$=0.30).
Filled symbols are used to show the models with a solar helium value. Error boxes correspond to Gonzalez et al. (\cite{gonzalez01}) (\textit{asterisks}), 
Santos et al. (\cite{santos04}) (\textit{diamonds}), Gillon \& Magain (\cite{gillon06}) (\textit{squares}), Fisher \& Valenti (\cite{fisher05}) (\textit{triangles}), 
and Ballot et al. (\cite{ballot11}) (\textit{crosses}) spectroscopic studies.}
   \label{edfitm182}
    \end{figure}

%%%%%%%%%%%%%%%%%%%%%%%%%%%%%%%%%%%%%%%%%%%%%%%%%%%%%%%%%%%%%%%%%%%%%%%%%%%%%%%%%%%%%%%%%%%%%%%%%%%%%%%%%%%%%%%%%%%%%%%%%%%%%%

\section{Scaling relations and automatic fits}

\subsection{Scaling relations}

The empirical scaling relations proposed by Kjeldsen \& Bedding (\cite{kjeldsen95}) can give approximate values of the mass and radius of a star from the observed average large separation, the frequency at the maximum of the power spectrum and the observed effective temperature:

\begin{equation}
    \frac{M}{M_{\sun}} = \left(\frac{135 \mu Hz}{<\Delta \nu>}\right)^{4} \left(\frac{\nu_{max}}{3050 \mu Hz}\right)^{3} \left(\frac{T_{eff}}{5777K}\right)^{3/2}
\end{equation}

\begin{equation}
    \frac{R}{R_{\sun}} = \left(\frac{135 \mu Hz}{<\Delta \nu>}\right)^{2} \left(\frac{\nu_{max}}{3050 \mu Hz}\right) \left(\frac{T_{eff}}{5777K}\right)^{1/2}
\end{equation}

For HD 52265, the frequency at the maximum amplitude is $\nu_{max}$ =  2090 $\pm$ 20 $\mu$Hz (Ballot et al. \cite{ballot11}). With a large separation of 98.4 $\mu$Hz and an effective 
temperature of 6100 K, we obtain from these relations $M = 1.23 M_{\sun}$ and $R = 1.32 R_{\sun}$. With a large separation corrected for the surface effects, of 99.4 $\mu$Hz, we obtain $M = 1.19 M_{\sun}$ and $R = 1.29 R_{\sun}$. In spite of the uncertainties, these results are in good agreement with our own results.

\subsection{Results from the SEEK code}

Computations have been done for this star using the SEEK automatic code (Quirion et al. \cite{quirion10}, and Gizon et al. \cite{gizon12}). 
This code makes use of 
a large grid of stellar models computed with the Aarhus Stellar Evolution Code (ASTEC). It searches for the best model corresponding to a seismically observed 
star, with the help of Bayesian statistics.The input parameters are the large and small average separations, the spectroscopic observables (T$_{eff}$, log g, [Fe/H]) 
and the absolute magnitude. The output gives the stellar mass, the radius and the age. In the case of HD 52265 the values given by the SEEK code for the mass and radius
are slightly larger than our results, and the age is smaller : $M = 1.27 \pm 0.03 M_{\sun}$, $R = 1.34 \pm 0.02 R_{\sun}$ and age = $2.37 \pm 0.29$ Gyr. 
The difference between the SEEK results and ours may be related to a different initial helium, to slightly different values of the average large and small separations as given by Gizon et al. (\cite{gizon12}), or to the fact that the SEEK results correspond to a secondary 
maximum of probability, as discussed below.

\subsection{Results from the Asteroseismic Modeling Portal}

We also performed computations for HD 52265 using the Asteroseismic Modeling Portal (https://amp.ucar.edu/). The AMP provides a web-based interface for
deriving stellar parameters for Sun-like stars from asteroseismic data. AMP was developed at the High Altitude Observatory and the Computational \& Information Systems Laboratory of the National Center for Atmospheric Research (Metcalfe et al. \cite{metcalfe09}). It uses the ASTEC and ADIPLS codes (Christensen-Dalsgaard 2008 a,b) coupled with a parallel genetic algorithm (Metcalfe \& Charbonneau \cite{metcalfe03}). Two different computations were done, the first one, AMP(a), by S. Vauclair using all the observed frequencies as given by Ballot et al. (\cite{ballot11}), and the second one, AMP(b), by S. Mathur using only the most reliable frequencies ((l=0, n=14), (l=1, n=14) and (l=2,n=15) were excluded). The final results are very close to the parameters found by using the TGEC code (Table 6).
Interestingly enough, the code found also solutions for a mass of 1.27 $M_{\sun}$ with a small Y (about 0.26) and a small age (about 2.7 Gyr) but
the $\chi^2$ tests showed that these results corresponded to secondary maxima, not to the best solution.

%%%%%%%%%%%%%%%%%%%%%%%%%%%%%%%%%%%%%%%%%%% TABLA final results TGEC %%%%%%%%%%%%%%%%%%%%%%%%%%%%%%%%%%%%%%%%%%%%%%%%

\begin{table}[h]
\caption{Final results for the parameters of the exoplanet-host star HD 52265 obtained with the TGEC code.}
\label{TGEC}
\centering
\begin{tabular}{ c c }     % 2 columns
\hline\hline

M/M$_{\sun}$ = 1.24 $\pm$ 0.02 & $[Fe/H]_i$ = 0.27 $\pm$ 0.04 \\
R/R$_{\sun}$ = 1.33 $\pm$ 0.02 & Y$_i$ = 0.28 $\pm$ 0.02 \\
L/L$_{\sun}$ = 2.23 $\pm$ 0.03 & $[Fe/H]_s$ = 0.20 $\pm$ 0.04 \\
log g = 4.284 $\pm$ 0.002 & Y$_s$ = 0.25 $\pm$ 0.02\\
Age (Gyr) = 2.6 $\pm$ 0.2 & T$_{eff}$ (K) = 6120 $\pm$ 20\\

\hline
\end{tabular}
\end{table}

%%%%%%%%%%%%%%%%%%%%%%%%%%%%%%%%%%%%%%%%%%%%%%%%%%%%%%%%%%%%%%%%%%%%%%%%%%%%%%%%%%%%%%%%%%%%%%%%%%%%%%%%%%%%%%%%

%%%%%%%%%%%%%%%%%%%%%%%%%%%%%%%%%%%%%%%%%%% TABLA AMP results %%%%%%%%%%%%%%%%%%%%%%%%%%%%%%%%%%%%%%%%%%%%%%%%

\begin{table}[h]
\caption{Final results for the parameters of the exoplanet-host star HD 52265 obtained from AMP automatic analysis, using all observed frequencies (a) or only the most reliable frequencies (b).}
\label{AMP}
\centering
\begin{tabular}{ c c c}     % 3 columns
\hline\hline

 & AMP(a) & AMP(b) \\
\hline

M/M$_{\sun}$ & 1.22  & 1.20 \\
R/R$_{\sun}$ & 1.321 & 1.310 \\
L/L$_{\sun}$ & 2.058 & 2.128 \\
log g & 4.282 & 4.282 \\      
$[Fe/H]$ & 0.23 & 0.215 \\
Y & 0.280 & 0.298 \\
Age (Gyr) & 3.00 & 2.38 \\
T$_{eff}$ & 6019 & 6097  \\

\hline
\end{tabular}
\end{table}

%%%%%%%%%%%%%%%%%%%%%%%%%%%%%%%%%%%%%%%%%%%%%%%%%%%%%%%%%%%%%%%%%%%%%%%%%%%%%%%%%%%%%%%%%%%%%%%%%%%%%%%%%%%%%%%%

\section{Summary and Conclusions}

In the present paper, we performed a detailed analysis of the exoplanet-host star HD 52265, which has been observed by CoRoT during 117 consecutive days, as one of the main targets. The beautiful observational results obtained for this star (Ballot et al. \cite{ballot11}) allowed a precise determination of its parameters, using classical comparisons between models computed with the TGEC and observational data. In our computations, we included atomic diffusion of helium and heavy elements. We found that, for the computed stellar models, the effects of radiative accelerations on individual elements is small and may be neglected. This result is consistent with the fact that detailed abundance analysis show similar enhancements in heavy elements compared to the Sun. We iterated the model computations so as to find a final surface [Fe/H]$_S$ in the observational range. We also compared these results with those obtained using approximate scaling relations (Kjeldsen \& Bedding \cite{kjeldsen95}), and automatic codes like SEEK (Quirion et al. \cite{quirion10}) and AMP (Metcalfe et al. \cite{metcalfe09}). Although the detailed physics included in the models is different, these results are in good agreement.

The good concordance between the results obtained with the TGEC code and the AMP for Sun-like stars was already proved with the star $\mu$ Arae. The results for this star, which were published separately (Soriano \& Vauclair \cite{soriano10} for TGEC; Do\u{g}an et al. \cite{dougan12} for AMP), are presented in the appendix for comparison. Altogether, these works represent real success of the asteroseismic studies for Sun-like stars.

\begin{acknowledgements} 

Much thanks are due to Travis Metcalfe for fruitful discussions and for introducing SV to AMP. Computational resources were provided 
by TeraGrid allocation TG-AST090107 through the Asteroseismic Modeling Portal (AMP). NCAR is partially supported by the Natiional Science Foundation.
We also thanks the referee for important and constructive remarks.

\end{acknowledgements}

\appendix

\section{}

Comparison of the results given by the TGEC analysis (Soriano \& Vauclair \cite{soriano08}) and the Asteroseismic Modeling Portal (Do\u{g}an et al.\cite{dougan12}) 
for the exoplanet-host sun-like star $\mu$ Arae.
%%%%%%%%%%%%%%%%%%%%%%%%%%%%%%%%%%%%%%%%%%% TABLA appendix mu Arae  %%%%%%%%%%%%%%%%%%%%%%%%%%%%%%%%%%%%%%%%%%%%%%%%
\begin{table}[h!]
\caption{Comparison of the results obtained with the TGEC and the Asteroseismic Modeling Portail for $\mu$ Arae.}
\label{AMP2}
\centering
\begin{tabular}{c c c}     % 3 columns
\hline\hline
 & TGEC & AMP \\
\hline

M/M$_{\sun}$ & 1.1 $\pm$ 0.02 & 1.1 \\
R/R$_{\sun}$ & 1.36 $\pm$ 0.06 & 1.365\\
L/L$_{\sun}$ & 1.90 $\pm$ 0.10 & 1.894 \\
log g & 4.215 $\pm$ 0.005 & 4.213 \\
T$_{eff}$ & 5820 $\pm$ 50 & 5807 \\
$[Fe/H]$ & +0.32 $\pm$ 0.02 & +0.418 \\
Y & 0.30 $\pm$ 0.01 & 0.316 \\
age(Gyr) & 6.34 $\pm$ 0.80 & 6.66\\

\hline
\end{tabular}
\end{table}

%%%%%%%%%%%%%%%%%%%%%%%%%%%%%%%%%%%%%%%%%%%%%%%%%%%%%%%%%%%%%%%%%%%%%%%%%%%%%%%%%%%%%%%%%%%%%%%%%%%%%%%%%%%%%%%%

\end{document}